# From Quantum Annealing to Alloy Discovery: Towards Accelerated Design of High-Entropy Alloys

Diego Ibarra-Hoyos[1]*, Peter Connors[2], Ho Jang[1], Nathan Grain[3], Israel Klich[1] , Gia-Wei Chern[1], Peter K. Liaw[3], John R. Scully[2,4] , Joseph Poon[1,2,*]

[1] Department of Physics, University of Virginia, Charlottesville, VA 22904, USA

[2] Department of Materials Science and Engineering, University of Virginia, Charlottesville, VA 22904, USA

[3] Department of Materials Science and Engineering, the University of Tennessee, Knoxville, TN 37996, USA

[4]Center for Electrochemical Science and Engineering, University of Virginia, Charlottesville, VA 22904, USA

*Corresponding author. Email: di8pd@virginia.edu (D.I.H.), sjp9x@virginia.edu (J.P.)

**ABSTRACT**

Data scarcity remains a central challenge in materials discovery, where finding meaningful descriptors and tuning models for generalization is critical but inherently a discrete optimization problem prone to multiple local minima confounding the true optimal state. Classical methods often get trapped in these minima, while quantum annealing can escape them via quantum fluctuations, including tunneling, that overcome narrow energy barriers. We present a quantum-assisted machine-learning (QaML) framework that employs quantum annealing to address these combinatorial optimization challenges through feature selection, support-vector training formulated in QUBO form for classification and regression, and a new QUBO-based neural-network pruning formulation. Recursive batching enables quantum annealing to handle large feature spaces beyond current qubit limits, while quantum-pruned networks exhibit superior generalization over classical methods, suggesting that quantum annealing preferentially samples flatter, more stable regions of the loss landscape. Applied to high-entropy alloys (HEAs), a data-limited but compositionally complex testbed, the framework integrates models for fracture-strain classification and yield-strength regression under physics-based constraints. The framework identified and experimentally validated Al8Cr38Fe50Mn2Ti2 (at.%), a single-phase BCC alloy exhibiting a 0.2 % yield strength of 568 MPa, greater than 40 % compressive strain without fracture, and a critical current density in reducing acid nearly an order of magnitude lower than 304 stainless steel. These results establish QA as a practical route to overcome classical optimization limits and accelerate materials discovery.

## INTRODUCTION

Optimization lies at the heart of scientific computing. From parameter fitting to model selection and adaptation, many problems reduce to finding configurations that minimize a cost function, a task that becomes exponentially difficult as solution spaces grow. Machine learning epitomizes this challenge: training neural networks, selecting features, and tuning hyperparameters all require navigating complex landscapes riddled with local minima. For problems of practical interest, classical gradient-based methods risk entrapment in suboptimal solutions[1–3], while exhaustive search remains computationally prohibitive. These challenges are particularly relevant to materials discovery, where machine-learning models increasingly rely on complex optimization to extract physical insight from limited and noisy data. The question of whether quantum computing can harness these challenges to accelerate such discrete optimization has thus emerged as a central thrust in near-term quantum algorithm development.

Quantum annealing (QA) offers a direct approach to this problem. By mapping optimization objectives into Hamiltonians, we transform discrete search problems into energy minimization tasks where the goal is to find the ground state, a natural fit for quantum systems [4,5]. QA exploits the quantum adiabatic theorem, which states that a quantum system remains in its instantaneous ground state if the Hamiltonian evolves slowly enough and maintains a finite energy gap from excited states. This principle enables QA to encode problems as Ising Hamiltonians and leverage quantum superposition and tunneling to explore exponentially large configuration spaces simultaneously, with the potential to escape local minima that trap classical solvers [4,5]. The mathematical bridge is the Quadratic Unconstrained Binary Optimization (QUBO) formulation, whose cost function maps directly onto an Ising Hamiltonian:

$$H = \sum_{i} h_i \sigma_i^z + \sum_{i<j} J_{ij} \sigma_i^z \sigma_j^z, \qquad 1$$

where spin variables $\sigma_i^z$ encode binary decisions and the ground state represents the optimal configuration. Crucially, this formulation is hardware-agnostic: QUBOs can be executed on current quantum annealers (e.g., D-Wave Advantage) or translated to gate-based architectures via the Quantum Approximate Optimization Algorithm (QAOA)[6,7], ensuring forward compatibility as quantum hardware matures. The practical question is therefore not whether quantum optimization offers theoretical advantages, but whether it can be harnessed to solve real world materials design challenges that classical methods struggle to address.

High-entropy alloys (HEAs) present precisely such a challenge. Unlike conventional alloys derived from one or two principal elements, HEAs combine four or more constituents in near-equiatomic ratios, producing high configurational entropy that stabilizes simple solid-solution phases instead of ordered intermetallics[8]. This compositional freedom opens a new path to attain the quest exceptional property combinations, strength, ductility, corrosion resistance, that knock down longstanding materials design trade-offs[8–21].

Yet rational discovery of such alloys remains severely constrained. The compositional space expands combinatorially with the number of elements[22], while traditional ab initio methods, though accurate, are too computationally intensive for systematic exploration. Heuristic design rules (e.g., atomic size mismatch, valence electron concentration)[23–25] capture qualitative trends but rarely predict multifunctional performance. Machine learning offers a scalable alternative, learning composition–property relationships directly from experimental data[22,26–37]. However, the HEA community faces an obstacle common to emerging materials systems: data scarcity. In this regime, conventional ML models face two interrelated difficulties: feature selection becomes critical to avoid overfitting in high-dimensional descriptor spaces, and model architectures must balance expressiveness against generalization. The optimization demands of these tasks are inherently combinatorial. QA provides a principled way to address such discrete optimization challenges[38–47] particularly relevant to feature selection, where only a few descriptors among hundreds are physically meaningful, and to neural-network pruning, where redundant neurons must be removed without sacrificing predictive accuracy.

Here we present a quantum-assisted machine-learning (QaML) framework that integrates feature selection, model training, and neural network pruning into a unified workflow for accelerated HEA discovery (Fig. 1).

HEA discovery of optimal alloys with precise compositions and phases to optimize multiple properties has not been realized due to the (i) significant change in properties indeed occur at abrupt thresholds brought about by small changes in compositions as well as structure and (ii) data scarcity.

This workflow is tailored to the data-scarce, high-noise regime characteristic of experimental HEA datasets, which are typically small and heterogeneous, collected under differing conditions that introduce statistical noise and bias. In such regimes, regression models can amplify noise rather than reveal physical trends. We therefore adopt a dual learning strategy: classification for fracture strain (ductile $\geq 30\%$ vs. brittle $< 30\%$), where measurements are particularly sensitive to microstructure and experimental fluctuations are high, and regression for yield strength, where data are more reproducible across studies.

At its core lie three algorithmic advances. First, we extend quantum support vector machines (QSVMs) from classification to regression by reformulating the loss function as a QUBO, enabling direct solution on quantum annealers, in a manner consistent with recent formulations of QUBO-based support vector regression[47,48]. This allows us to address both fracture-strain classification and yield-strength prediction within a unified quantum optimization framework. Second, we introduce a novel quantum-assisted neural network pruning formulation for fully connected architectures that preserves downstream activation dependencies, allowing entire neuron subsets to be optimized discretely while maintaining functional fidelity. While earlier studies have explored quantum-based pruning in convolutional settings[49], our formulation targets dense feedforward networks, addressing the small-data scientific regime where such architectures are most commonly employed. Third, we implement quantum-based feature selection (QBoost and Quantum Mutual Information) with a recursive batching strategy that scales to high-dimensional feature spaces despite current qubit limitations.

We integrate these components into a materials-discovery pipeline targeting single-phase BCC alloys in the Al–Cr–Fe–Mn–Ti system. Corrosion resistance was incorporated as a design constraint through the inclusion of Cr, Al, and Ti as passivating elements[50–55], along with upper bounds on Al and Mn to mitigate oxide instability and pitting[56–58]. These corrosion-conscious limits, together with density ($<7$ g cm$^{-3}$), phase-stability, and cost criteria (excluding expensive refractory elements such as Nb, Hf, and Zr), defined the design space explored by QaML. Within this constrained space, multi-objective screening identified

Al8Cr38Fe50Mn2Ti2 (at.%) as the optimal candidate. Experimental synthesis confirmed the predicted phase stability and mechanical performance, while corrosion behavior was subsequently benchmarked against 304 stainless steel to validate the corrosion-aware design constraints. This outcome demonstrates that quantum annealing, coupled with machine-learning models under physics-informed constraints, can guide materials discovery from computational prediction to experimental realization, establishing QaML as a practical framework for navigating complex design spaces where classical optimization alone proves insufficient.

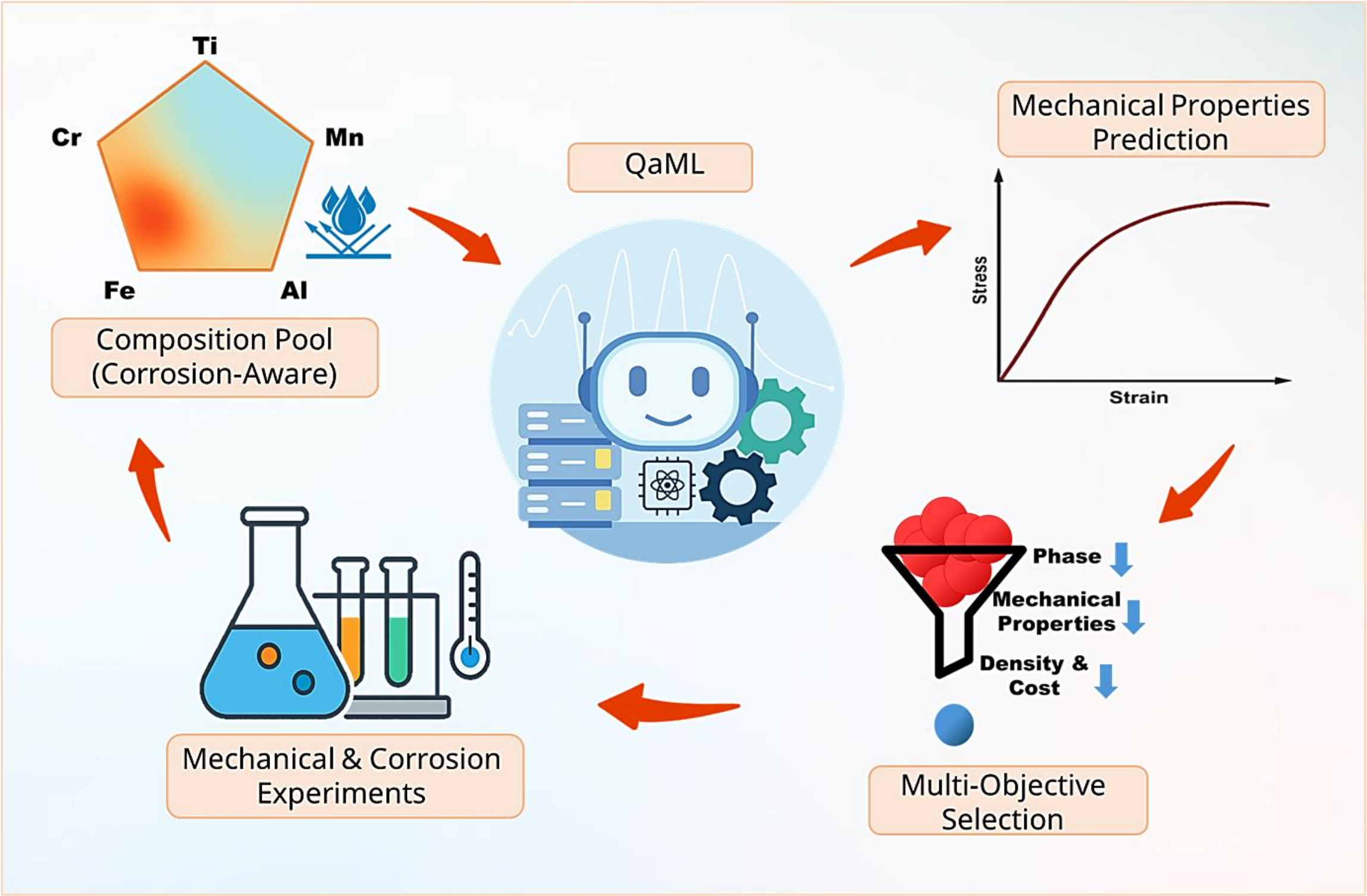

**Figure 1. Quantum-enhanced workflow for the multi-objective discovery of high-entropy alloys.** The integrated pipeline combines quantum-assisted machine-learning models with physics-based constraints to guide alloy design. Starting from a corrosion-aware composition pool within the Al–Cr–Fe–Mn–Ti system (top left), quantum-assisted machine learning (QaML) models predict key mechanical properties such as yield strength and fracture strain (top right). A sequential screening cascade (bottom right) evaluates the phase stability, ductility, density, and cost to down-select viable candidates within the Al–Cr–Fe–Mn–Ti design space. The most promising composition, Al8Cr38Fe50Mn2Ti2, emerges as the optimized solution and is subsequently synthesized and characterized experimentally for mechanical and corrosion performance (bottom left), demonstrating purposeful materials discovery driven by quantum-machine-learning inference.

The remainder of this paper is organized as follows. We first describe quantum-assisted feature selection via QBoost[59–62] and Mutual Information [63], benchmarking QA against classical simulated annealing. We then present QSVM formulations for both classification and regression, followed by our novel neural network pruning framework. Finally, we deploy optimized models in a systematic screening workflow, culminating in the experimental validation of Al8Cr38Fe50Mn2Ti2 as a high-performance BCC alloy meeting simultaneously targeted strength, ductility, corrosion, density, and cost objectives. This work establishes a template for quantum-enhanced materials discovery: formulating discrete design decisions as QUBOs, solving them on near-term quantum hardware (D-Wave Advantage), and validating predictions experimentally, bridging the gap between quantum algorithmic development and tangible materials innovation.

## RESULTS

### Scalable Quantum Feature Selection

Feature selection defines the foundation of the quantum-assisted machine learning (QaML) workflow, determining which physical descriptors most effectively capture structure–property relationships. While predictive performance remains the primary objective, constraining the feature space through well-structured subsets also promotes interpretability by ensuring that the resulting models retain physically meaningful descriptors without further post-processing. The feature library, totaling 255 variables, includes thermodynamic and electronic descriptors (e.g., mixing enthalpy, configurational entropy, atomic-size mismatch, and elastic moduli)[37,64–68] together with elemental compositions, ratios, and sums (full list in Supplementary Information Section A).

To construct compact and physics-consistent subsets, we combined three complementary selectors: Lasso, QBoost[59–62], and Quantum Mutual Information (Q-MI)[63]. Lasso imposes linear sparsity through an $\ell_1$ penalty; QBoost identifies features that strengthen margin-based decision ensembles; and Q-MI balances feature–target relevance against inter-feature redundancy (QUBO formulations detailed in Methods). Each introduces a distinct selection bias, allowing us to probe different facets of the compositional design space.

A key enabler of scalability is the recursive batching strategy, which partitions the feature library into smaller sub-QUBOs, solves each batch independently on the D-Wave system, and pools the retained features across iterations (see Methods). Beyond overcoming hardware limits, this design deliberately encourages diversity by exploring multiple low-energy regions of the search space rather than converging to a single basin, consistent with the multi-solution sampling advantages[69].

We benchmarked quantum annealing (QA) against classical simulated annealing (SA) across batch sizes $K = 20, 50, 100$ and multiple hyperparameter configurations. For QBoost, both solvers consistently converged to identical feature subsets, indicating that the optimization landscape is relatively smooth. However, QA achieved these solutions with lower and more stable runtimes (Fig. 2, right), which scaled gently with batch size, demonstrating a present-day, practical advantage in speed and consistency even on existing hardware.

In contrast, Q-MI presented a much rougher energy landscape, sensitive to the redundancy ($\lambda_1$) and sparsity ($\lambda_2$) parameters (see Methods). Here, the distinction between QA and SA became pronounced: across parameter sweeps and batch sizes, QA more frequently identified lower-energy configurations corresponding to superior feature subsets (Table S2). This behavior reflects quantum annealing's capacity to traverse narrow barriers through tunneling and reach minima that classical SA often misses.

These results establish QA as a reliable optimization tool across contrasting feature-selection landscapes. For smoother objectives such as QBoost, QA matches classical performance but delivers faster, stable scaling with problem size. For more rugged formulations like Q-MI, QA accesses better minima without increased computational cost, maintaining the same runtime efficiency observed in QBoost. The recursive batching strategy extended these advantages to large descriptor sets, maintaining stable scaling and promoting diverse solution exploration. These optimized feature subsets served as the foundation for subsequent model training, where their impact on predictive performance and generalization was systematically evaluated across classical and quantum-enhanced architectures.

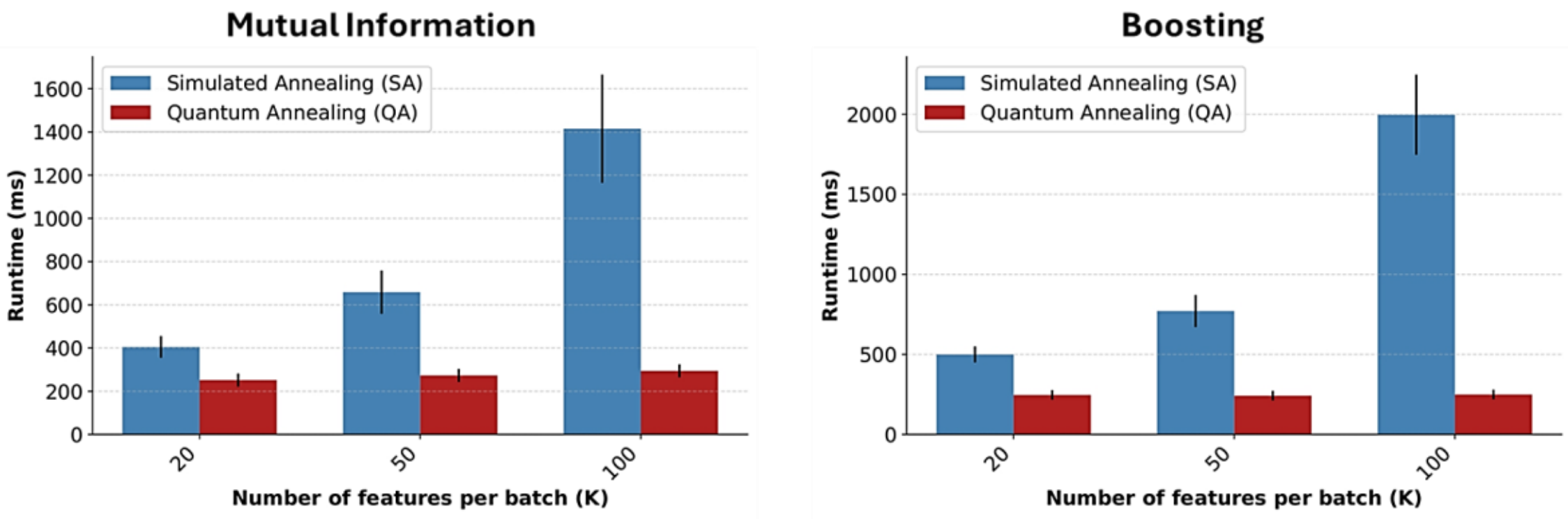

**Figure 2. Runtime comparison of Simulated Annealing (SA) and Quantum Annealing (QA) for QUBO-based feature selection.** Average runtime per QUBO instance as a function of problem size, expressed as the number of features per batch (K =20,50,100) for (left) Mutual Information classification and (right) Boosting (classification). SA runtimes (blue) increase sharply with problem size, while QA runtimes (red) remain comparatively stable. Note that these measurements correspond to the solution of a single QUBO problem rather than the total runtime of the full recursive batching workflow.

## Model Selection

To evaluate how well each feature subset enables property prediction, we trained four machine learning architectures: Random Forests (RF), Support Vector Machines (SVM), Quantum-assisted Support Vector Machines (QSVM) and Neural Networks (NN), using 5-fold cross-validation. For fracture-strain classification, we assessed performance through accuracy and F1 scores; for yield-strength regression, we used Root Mean Square Error (RMSE) and $R^2$. The best-performing feature sets identified during cross-validation were then retrained on the full training data and tested on a held-out dataset to confirm generalization (Table 1). Complete implementation details appear in Methods.

The results reveal an alignment between feature selection strategy and optimal model architecture. For fracture-strain classification, QBoost-selected features consistently delivered superior performance across all models, with Random Forest achieving the highest training accuracy. This synergy makes physical sense: both QBoost and RF build predictions from tree-based decision boundaries, creating a natural compatibility that enhances both performance and interpretability. However, the models exhibiting the best generalization, SVM and QSVM, also relied on QBoost-selected features, indicating that while the RF–QBoost pairing captures the training structure most effectively, margin-based architectures benefited more from its generalizable feature subsets. The performance and formulation of QSVM are examined in greater detail in the following subsection. For yield-strength regression, Q-MI feature subsets proved most effective, though optimal performance depended on carefully tuning the redundancy–sparsity balance (λ). SVM and QSVM presented an interesting exception, owing to its linear kernel, it performed best with Lasso-selected features; a result that reflects the fundamental alignment between Lasso's linear sparsity enforcement and SVM's linear margin optimization. (The final feature subsets corresponding to these best-performing models are listed in Table S1.)

**Table 1 | Model-selection results under cross-validation and test evaluation.** (a) Classification of fracture strain using QBoost-selected features. Reported metrics are accuracy and F1 for Random Forest (RF), Support Vector Machine (SVM), Quantum enhance SVM (QSVM), and Neural Network (NN). (b) Regression of yield strength using Q-MI or Lasso-selected features, with RMSE and $R^2$reported for RF, SVM, QSVM, and NN. Training metrics reflect 5-fold cross-validation; test metrics reflect evaluation on the held-out test set. In the Q-MI formulation (Method Section), $\lambda_1$ and $\lambda_2$ act as trade-off parameters penalizing feature redundancy and subset size, respectively.

(a) Classification (fracture strain)

| **Algorithm** | **Model** | **Accuracy (Train)** | **F1 (Train)** | **Accuracy (Test)** | **F1 (Test)** |
|---|---|---|---|---|---|
| RF | QBoost | 1 | 1 | 0.821 | 0.824 |
| SVM | QBoost | 0.903 | 0.901 | 0.897 | 0.899 |
| QSVM | QBoost | 0.845 | 0.762 | 0.897 | 0.899 |
| NN | QBoost | 0.981 | 0.971 | 0.872 | 0.815 |

(b) Regression (yield strength)

| **Algorithm** | **Model** | **RMSE (Train)** | **$R^2$ (Train)** | **RMSE (Test)** | **$R^2$ (Test)** |
|---|---|---|---|---|---|
| RF | Q-MI ($\lambda_1 = 0.1,\ \lambda_2 = 0.01$) | 70.61 | 0.981 | 274.49 | 0.72 |
| SVM | LASSO | 212.61 | 0.83 | 373.41 | 0.48 |
| QSVM | LASSO | 305.90 | 0.64 | 349.13 | 0.55 |
| NN | Q-MI ($\lambda_1 = 0.1,\ \lambda_2 = 0.001$) | 184.28 | 0.87 | 325.03 | 0.61 |

## Quantum-Assited Support Vector Machines

Quantum-assisted support vector machines (QSVMs) were implemented for both the fracture-strain classification and yield-strength regression. For classification, the formulation followed the quantum SVM framework of Willsch et al.[47] and the implementation protocol established in our previous QBoost study[62]. The classical dual problem,

$$L = \frac{1}{2}\sum_{i,j} \alpha_i \alpha_j y_i y_j \phi(x_i, x_j) - \sum_i \alpha_i \qquad 2$$

subject to $0 \leq \alpha_i \leq C$ and $\sum_i \alpha_i y_i = 0$, was mapped to a QUBO representation for annealing-based optimization (see Methods).

For regression, the QUBO formulation is derived by adapting the dual loss function[48]

$$L = \frac{1}{2}\sum_{i,j} (\alpha_i - \alpha_i^*)(\alpha_j - \alpha_j^*)\phi(x_i, x_j) + \varepsilon \sum_i (\alpha_i - \alpha_i^*) - \sum_i y_i(\alpha_i - \alpha_i^*) \qquad 3$$

subject to $0 \leq \alpha_i, \alpha_i^* \leq C$ and $\sum_i (\alpha_i - \alpha_i^*) = 0$. (Derivation in Methods and Supplementary Information).

Unlike the classification case, regression requires encoding both dual parameter sets $\alpha$ and $\alpha^*$, effectively doubling the number of binary variables and substantially increasing embedding complexity due to the limited hardware connectivity. This qubit overhead represents a principal limitation of the present hardware generation.

To alleviate these constraints in both tasks, particularly in regression, two strategies were employed: consistent data splits and preprocessing following Willsch et al.[47], and the use of D-Wave's Hybrid Leap solver. These approaches effectively reduced embedding overhead and enabled convergence within the available qubit resources. Among them, the Hybrid Leap solver yielded the most stable performance. Full implementation details are provided in the Supplementary Information Sections D & E.

For regression, the QSVM demonstrated improved generalization compared to the classical SVM, likely due to the relaxed constraint formulation inherent in the QUBO encoding rather than any direct quantum advantage, as discussed in our previous study[62]. In contrast, for classification, both quantum and classical SVMs exhibited comparable performance (Table 1).

## Quantum-Enhanced Neural-Network Pruning

Neural networks trained on limited datasets face a fundamental challenge: balancing model complexity against overfitting risk. We addressed this trend by reformulating neuron pruning as a discrete optimization problem directly solvable by the quantum annealing hardware. The approach casts pruning decisions as a Quadratic Unconstrained Binary Optimization (QUBO) problem, where each binary variable $\beta_i$, determines whether neuron $i_{th}$ is retained, with $k$ specifying the target network size.

$$\min_{\beta\epsilon\{0,1\}^n} \left\|\mathbf{A}^{[l]} - \mathbf{A}'^{[l]}(\boldsymbol{\beta})\right\|_F^2 + \mu\left\|\mathbf{Z}^{[l+1]} - \mathbf{Z}'^{[l+1]}(\boldsymbol{\beta})\right\|_F^2 + \lambda\left(\sum_i \beta_i - k\right)^2 \quad 4$$

The optimization objective balances three competing demands: reconstruction of layer activations, preservation of downstream pre-activations, and a quadratic sparsity constraint. Simply removing neurons with the weakest activations, equivalent to minimizing $\left\|\mathbf{A}^{[l]} - \mathbf{A}'^{[l]}(\boldsymbol{\beta})\right\|_F^2$, risks propagating errors through subsequent layers (Fig. S1). Because downstream pre-activations depend on the pruned layer through $\mathbf{Z}^{[l+1]} = \mathbf{A}^{[l]}\mathbf{W}^{[l+1]} + b^{[l+1]}\mathbf{1}^T$, even weak neurons can have substantial network-wide influence. Our formulation therefore includes a second term to preserve these downstream effects, while a quadratic penalty enforces the desired sparsity level. This process yields a fully quadratic objective suitable for direct quantum annealing implementation. Though applicable to any number of layers, we demonstrate the approach on single hidden layers within our compact network architectures, with full technical details in Methods. A schematic illustration of the pruning mechanism is provided in Fig. 3.

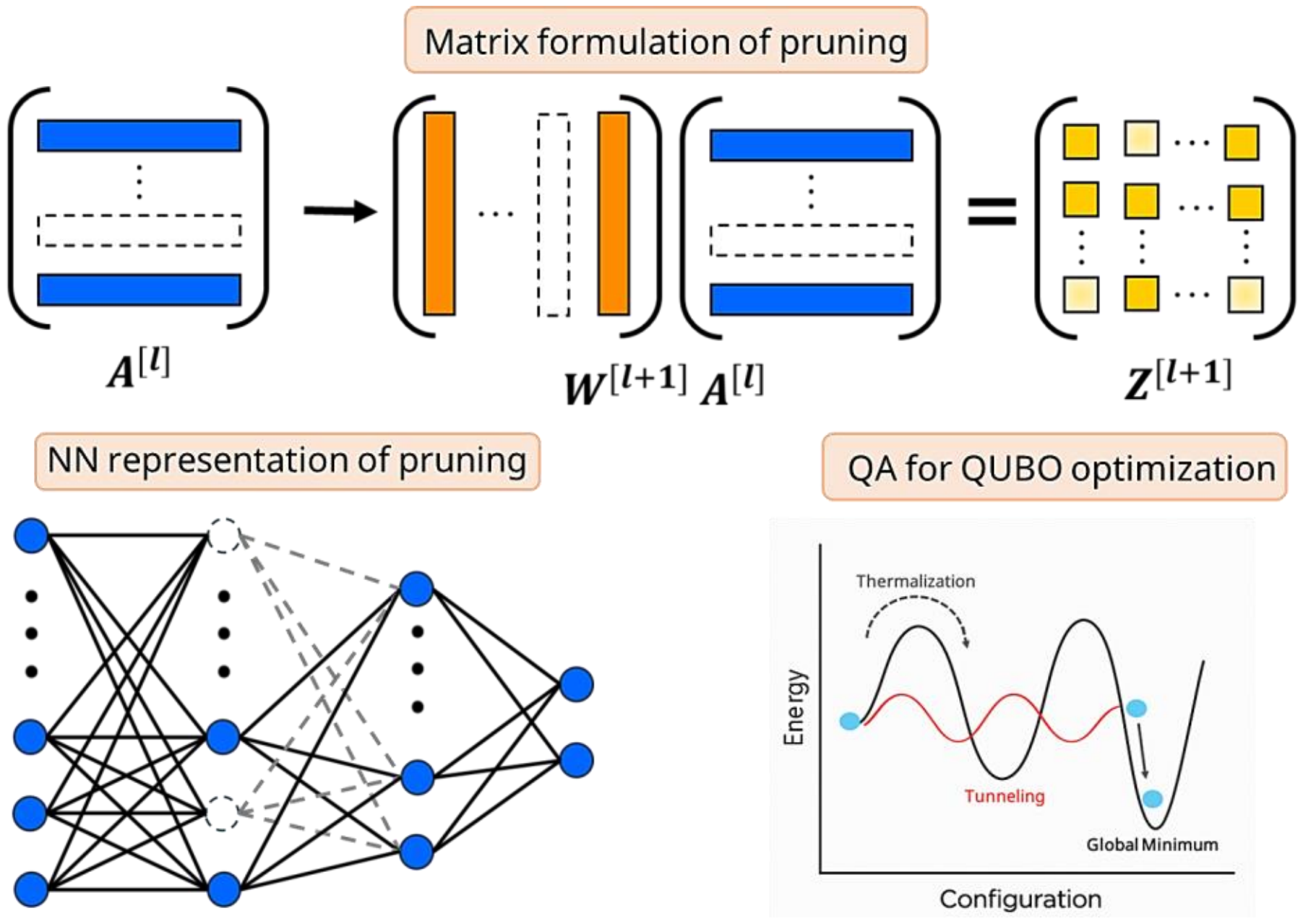

**Figure 3. Schematic illustration of quantum-annealing–based neural network pruning.** Matrix formulation of pruning, where selected columns corresponding to pruned neurons are removed. For clarity, the bias term is not shown in the scheme, as it is unaffected by pruning, although it is included in the full formulation. Neural-network representation, with pruned neurons and their outgoing connections, greyed out. In the quantum-annealing schematic, tunneling allows the exploration beyond narrow minima to identify broader, more generalizable pruning configurations.

The results reveal an advantage of quantum annealing. For fracture-strain classification, we pruned a (32, 32, 16) network by removing 16 neurons from the second hidden layer. The quantum-pruned network outperformed every alternative tested, including the Random Forest, Support Vector Machines, the original unpruned network, and an identically pruned network optimized by simulated annealing (Table 2). Most remarkably, simulated annealing consistently achieved lower QUBO energies (Table S3), suggesting better optimization convergence, yet the quantum-annealed solutions generalized more effectively to test data. This behavior indicates that QA may be sampling qualitatively different regions of the optimization landscape, a point explored in detail in the Discussion.

For yield-strength regression, we aggressively pruned a (64, 64, 32) network by eliminating 30 of 64 neurons in the first hidden layer. The quantum-pruned model maintained performance statistically equivalent to both the full network and its classically pruned counterpart (Table 2). While this case did not show the generalization boost observed in classification, achieving this size reduction without accuracy loss demonstrates a clear value for resource-constrained deployment scenarios.

**Table 2 | Performance comparison of baseline and pruned neural networks.** (a) Fracture strain classification and (b) yield strength regression. Reported metrics are the accuracy and F1 score for classification, and RMSE and $R^2$ for regression. "Baseline NN" refers to the unpruned network, while the "QA-pruned NN" and "SA-pruned NN" denote models pruned using quantum annealing and simulated annealing, respectively. In both cases pruning was applied to a single hidden layer as specified in the text. For clarity, only test-set values are shown for pruned models; training metrics for the baseline are reported for reference.

(a) Classification (fracture strain)

| **Algorithm** | **Accuracy (Train)** | **F1 (Train)** | **Accuracy (Test)** | **F1 (Test)** |
|---|---|---|---|---|
| Baseline NN | 0.981 | 0.971 | 0.872 | 0.815 |
| QA-pruned NN | - | - | 0.923 | 0.889 |
| SA-pruned NN | - | - | 0.872 | 0.815 |

(b) Regression (yield strength)

| **Algorithm** | **RMSE (Train)** | **$R^2$ (Train)** | **RMSE (Test)** | **$R^2$ (Test)** |
|---|---|---|---|---|
| Baseline NN | 184.276 | 0.873 | 325.032 | 0.61 |
| QA-pruned NN | - | - | 323.62 | 0.61 |
| SA-pruned NN | - | - | 321.126 | 0.62 |

## Accelerated Alloy Discovery Through Integrated Screening

Having established optimized predictive models, we deployed them in a systematic screening workflow designed to identify high-performance alloys meeting multiple competing design criteria. Each mechanical property was predicted using its best-performing model: the quantum-pruned neural network for fracture-strain classification (binary threshold at 30% ductility) and the Random Forest regressor for yield-strength. To ensure practical viability and corrosion resistance, we constrained the search space to single-phase body-centered-cubic (BCC) alloys, with phase stability verified through both our ML-based classifier and independent CALPHAD thermodynamic calculations.

We focused the compositional search on the Al–Cr–Fe–Mn–Ti system, with Cr and Ti primary passivators[52–55] and Al as secondary passivator[52,53], while deliberately excluding heavier, high-cost elements like Nb, Zr, and Hf. These elements were selected for their synergistic roles in passivation. Upper bounds on Al and Mn were imposed to suppress oxide heterogeneity and prevent the formation of Mn-rich, non-protective oxides[56–58]. Table 3 summarizes the screening ranges, with the aluminum composition determined by balance.

For each candidate composition within this constrained design space, we simultaneously evaluated phase stability, fracture strain classification, yield strength, density, and relative cost. Subsequent ranking emphasized physical feasibility over purely statistical optimization: single-phase BCC stability was treated as a hard constraint, followed by the requirement of fracture strain ≥ 30% as predicted by the pruned neural network classifier. Among the compositions meeting these two criteria, candidates were further prioritized by low density and acceptable cost (≤ \$3.00 $kg^{-1}$) while yield strength was used as a secondary indicator of mechanical competitiveness rather than a limiting factor.

**Table 3 |** Screening ranges (at.%) and step sizes for Cr, Fe, Mn, and Ti in the Al–Cr–Fe–Mn–Ti system, with Al taken as balance to maintain 100 at.%.

| Al | Cr | Fe | Mn | Ti |
|---|---|---|---|---|
| bal | [15-40] step: 1 | [40-60] step:5 | [2-10] step:1 | [2-5] step:1 |

This multi-stage workflow identified several compositions satisfying all design constraints. Among these, Al8Cr38Fe50Mn2Ti2 (at.%) was selected as the optimal candidate due to its lowest density (6.96 g $cm^{-3}$) while maintaining a predicted yield strength of 990 MPa, placing it in the high-strength regime. CALPHAD analysis confirmed its single-BCC structure (see Fig. S6). Its combination of high ductility, low density, and competitive raw-material cost of \$2.20 $kg^{-1}$ [70,71] makes it attractive for weight-sensitive and cost-constrained structural applications. This outcome illustrates how physics-informed constraints coupled with data-driven screening can efficiently guide alloy discovery within vast compositional spaces toward experimentally realizable targets.

We synthesized Al8Cr38Fe50Mn2Ti2 and characterized its phase constitution and mechanical response. Although corrosion resistance was not explicitly modeled, its consideration during composition selection, through the inclusion of Cr, Al, and Ti as synergistic passivators, motivated direct benchmarking against 304 stainless steel to assess chemical durability alongside mechanical performance. These experimental validations are presented in the following sections.

## Mechanical Testing

$Al_8Cr_{38}Fe_{50}Mn_2Ti_2$ alloys were synthesized via arc melting and subjected it to comprehensive structural and mechanical characterization. X-ray diffraction (XRD) confirms the alloy crystallizes as a single-phase body-centered cubic solid solution, with only the characteristic (110) and (200) reflections present and no detectable secondary phases or intermetallics (Fig. 4A). This finding validates both the machine learning phase classifier and the CALPHAD thermodynamic predictions.

To confirm compositional homogeneity, energy dispersive spectroscopy (EDS) area maps were collected across the surface of $Al_8Cr_{38}Fe_{50}Mn_2Ti_2$ (Fig. 4C). A representative backscattered electron (BSE) image and corresponding EDS map are shown in Fig. 4. Despite the presence of some porosity in the BSE micrograph, no elemental segregation or inhomogeneity was detected, corroborating the XRD evidence of a single-phase microstructure.

Uniaxial compression testing reveals a 0.2% offset yield strength of 568 MPa and engineering strains exceeding 40% without fracture (Fig. 4B), confirming the quantum-pruned neural network's prediction of ductile behavior (>30% threshold). The continuous work-hardening response drives true stress beyond 2.3 GPa (1.5 GPa engineering stress) within the accessible strain window, with the specimen remaining intact at test termination. These values represent the maximum measured stresses rather than an ultimate strength, as the sample did not fracture.

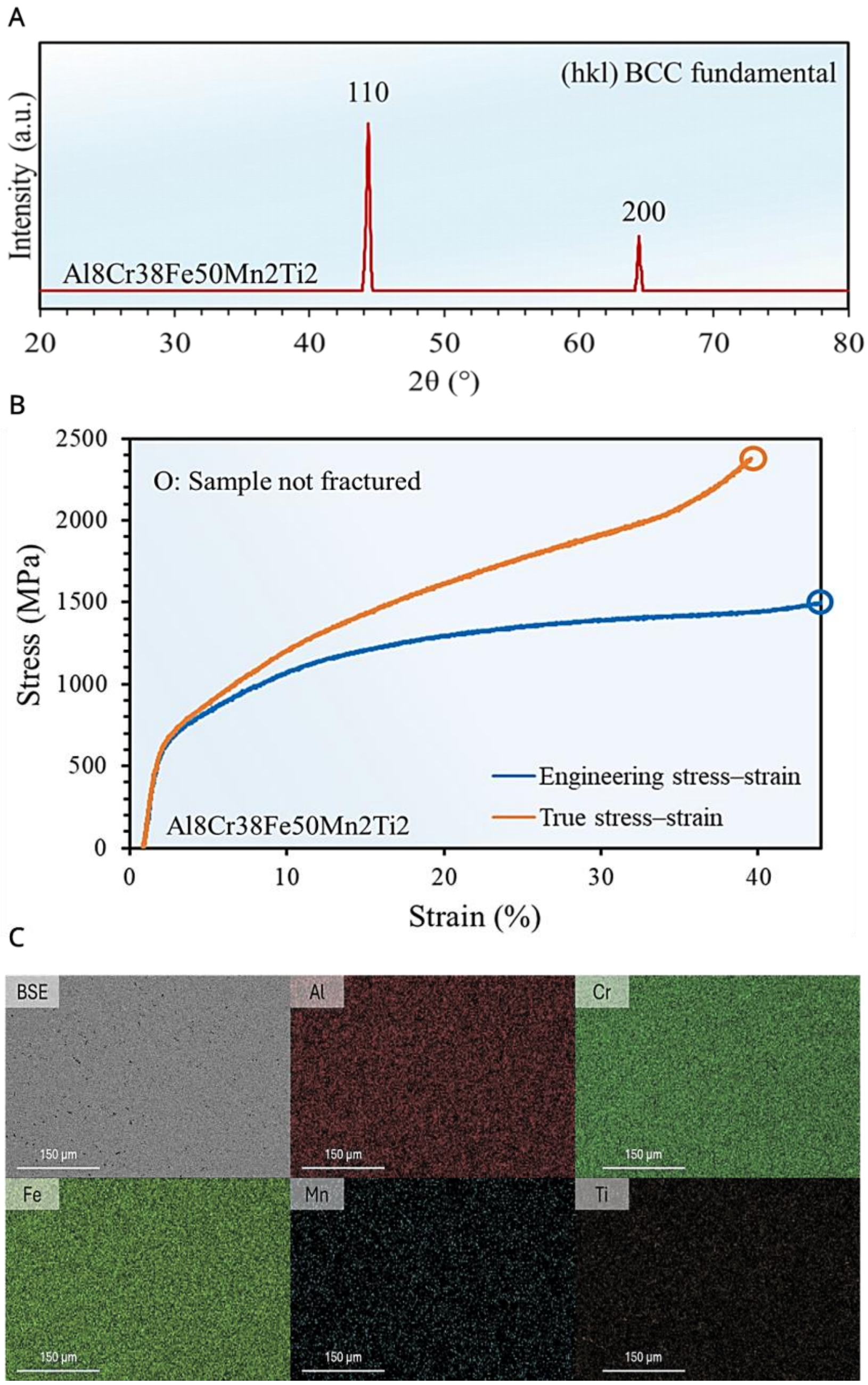

**Figure 4. Phase and mechanical validation of the selected Al8Cr38Fe50MnnTi2 alloy**. (a) XRD pattern showing peaks corresponding exclusively to the fundamental BCC reflections (110, 200), confirming a single-phase BCC solid solution with no detectable secondary intermetallic. (b) Representative engineering and true stress–strain curves obtained under uniaxial compression, indicating a 0.2% offset yield strength of 568 MPa, an engineering fracture strain of 40%, with no evidence of failure within the measured strain window. True stress and true strain values were calculated from the engineering values using $\sigma_t = \sigma_e(1 + \varepsilon_e)$ and $\varepsilon_t = \ln(1 + \varepsilon_e)$ where (c) EDS area maps across the surface of the Al8Cr38Fe50Mn2Ti2.

## Corrosion Testing

In order to demonstrate the enhanced corrosion resistance and passivity of the selected Al8Cr38Fe50MnTi2 alloy, open circuit potential (OCP) and potentiodynamic polarization (PD) experiments were performed on both Al8Cr38Fe50Mn2Ti2 and 304 grade stainless steel, (containing 8 wt.% Ni and 18 wt.% Cr), an industrially relevant stainless steel, in two solutions of increasing harshness. In this method, the relative passivity is assessed by increased OCP and decreasing or fixed anodic current density with increasing potential driving force.

OCP monitoring was performed for 15 minutes before PD, and representative OCP E-t curves are presented in Fig. 5A. The OCP of both Al8Cr38Fe50Mn2Ti2 and 304 stainless steel decrease between solutions of increasing harshness (i.e. higher [$Cl^-$]), matching expected trends. Notably, Al8Cr38Fe50Mn2Ti2 is demonstrated to have a higher terminal OCP than 304 stainless steel in either solution, further highlighting the observed benefits of including passivating elements such as Cr, Ti and Al in the CCA relative to the stalwart 304[50–53].

The resulting E-log(*i*) curves from the PD experiments performed following OCP are presented in Fig. 5B Considering critical current density ($i_{crit}$), the expected result that increasing solution harshness increases $i_{crit}$ is observed for each sample across the different solutions. However, the decrease in $i_{crit}$ indicates improved passivation at high potential driving force. Across samples, Al8Cr38Fe50Mn2Ti2 exhibits a nearly order of magnitude decrease in critical current density relative to 304 stainless steel, indicating enhanced passivation efficiency in either solution. Furthermore, passive current densities ($i_{pass}$) in either solution are shown to be less for Al8Cr38Fe50Mn2Ti2 than for 304 stainless steel, indicating a more protective passive film being grown on Al8Cr38Fe50Mn2Ti2 relative to 304 stainless steel in both solutions.. Finally, Al8Cr38Fe50Mn2Ti2 reflects remarkable pitting resistance in either solution, resisting pitting corrosion before transpassive dissolution occurs (typically at about 1.2 $V_{SHE}$ for $Cr_2O_3$-based passive films). In comparison, 304 stainless steel's passive film failed by pitting at about 0.5 $V_{SHE}$ and 0.8 $V_{SHE}$ in the 0.1 M H2SO4+0.5 M NaCl and 0.1 M H2SO4+0.25 M NaCl solutions, respectively, indicating a relatively less stable passive film. In summary, Al8Cr38Fe50Mn2Ti2 is demonstrated to have markedly improved corrosion resistance to 304 stainless steel in either solution, achieving a higher OCP, lower $i_{crit}$ and $i_{pass}$, and enhanced passive film stability.

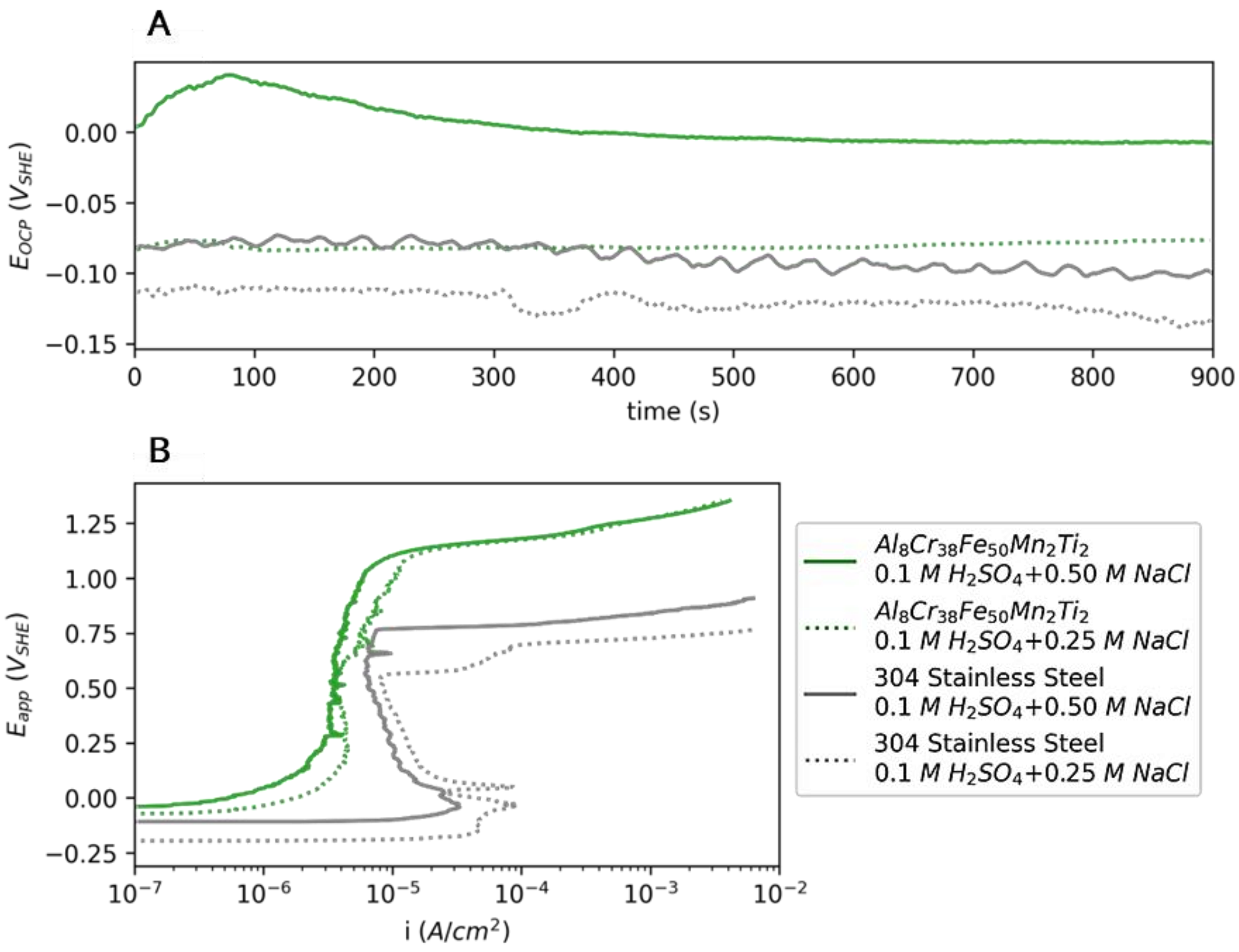

**Figure 5. Open-circuit potential (OCP) and potentiodynamic polarization (PD) results of the selected Al8Cr38Fe50MnTi2 alloy and 304 grade stainless steel**. (a) Representative OCP-monitoring curves for both samples in either solution, performed over 15 minutes. (b) Representative E-log(*i*) curves resulting from PD experiments performed on both samples in each solution. In all experiments, solution was constantly deaerated using $N_2$ gas bubbled into the input solution reservoir into the SDC, and n = 5 point boxcar averaging was utilized to decrease the signal to noise ratio in presented curves.

## DISCUSSION

This work demonstrates how quantum annealing can be integrated into a coherent, solver-agnostic framework for materials discovery. The QaML architecture developed here unifies feature selection, model optimization, and neural-network pruning under a common QUBO formalism, bridging algorithmic advances in quantum optimization with the practical constraints of small, noisy materials datasets. Beyond achieving competitive predictive accuracy, the results highlight how each quantum component contributes distinct value to the overall design workflow.

Feature selection emerged as inherently property-dependent, reflecting differences in the physics underlying ductility and strength. For fracture-strain classification, margin-oriented QBoost identified descriptors associated with microstructural stability, while for yield-strength regression, Q-Mutual Information emphasized features linked to lattice-distortion and solid-solution mechanisms (Fig. 6)[72,73]. Although the QaML models were designed for predictive optimization rather than mechanistic inference, the selected descriptors retained clear physical meaning (Fig. 6), illustrating that the quantum-optimized models remain interpretable and grounded in materials physics. Recursive batching proved essential for scalability, enabling high-dimensional feature exploration on limited-connectivity hardware without sacrificing performance or diversity of solutions.

A persistent challenge lies in embedding the logical QUBO graph into the physical qubit topology of the D-Wave processor (Fig. S4, S5). Current heuristic minor-embedding algorithms impose a substantial overhead, effectively limiting the number of usable qubits and thus the maximum problem size. Even within these limits, QA maintained stable performance across all batch sizes and consistently matched or outperformed simulated annealing (SA) in solution quality. The recursive batching strategy was particularly beneficial: solving smaller QUBOs independently promoted exploration of multiple low-energy basins rather than converging to a single global minimum, analogous to the diversity benefits of mini-batch training in neural networks[69]. Preliminary benchmarks further showed that QA runtime scaled more gently with problem size than classical SA, offering up to an order-of-magnitude advantage for the largest batches tested, highlighting its promise as a consistent and scalable optimization primitive even under current hardware constraints.

Although QSVMs were not the top-performing models, they demonstrated generalization comparable to classical SVMs. This behavior suggests that the QUBO reformulation of the dual problem effectively smooths the constraint landscape, allowing the optimizer to explore near-feasible regions that classical solvers typically exclude. The relaxed constraint structure may thus serve as a form of implicit regularization, producing models that generalize well even without achieving lower training error.

The QUBO-based neural-network pruning represents the first application of QA to model compression within small-data materials science. For fracture-strain classification, QA-pruned networks surpassed both their unpruned baselines and SA-pruned counterparts despite converging to slightly higher QUBO energies. This counterintuitive outcome aligns with recent theoretical and experimental studies showing that quantum annealing preferentially samples broad, high-entropy basins of the energy landscape rather than narrow isolated minima[74–76]. Quantum fluctuations effectively delocalize the search across dense clusters of near-degenerate low-energy states (Fig. S3), lowering their effective energy and biasing the process toward flat regions associated with better generalization[69,77–79]. In this sense, QA acts as an implicit regularizer. Rather than driving the network to the absolute minimum, it finds configurations that are more stable under perturbations and therefore generalize better to unseen data. For yield-strength regression, QA and SA achieved comparable results while substantially reducing network size, confirming the framework's efficiency even in the absence of a generalization advantage.

The integration of quantum optimization into the HEA design loop demonstrates a direct path from algorithmic innovation to experimental validation. The QaML-driven screening identified Al8Cr38Fe50Mn2Ti2 (at.%) as an optimal single-phase BCC alloy balancing strength and ductility, with corrosion resistance emerging naturally from the physics-based composition constraints defining the search space. Experimental synthesis confirmed a 0.2 % yield strength of 568 MPa and >40 % compressive strain without fracture, alongside broad improvements in corrosion resistance, even achieving a nearly order-of-magnitude decrease in $i_{crit}$ relative to the stalwart 304 stainless steel. Its combination of low density (~6.96 g $cm^{-3}$ compared to ~7.79 g $cm^{-3}$ for 304 stainless-steel), moderate cost (~$2.20 $kg^{-1}$, 304 stainless-steel reference ~$1.72 $kg^{-1}$, subject to price fluctuations) [70,71], and superior corrosion resistance highlights its potential for marine and weight-sensitive structural applications. This convergence between computational prediction and experimental validation underscores the practical relevance of quantum-enhanced optimization in real materials systems.

Taken together, these results move quantum annealing from an imaginative concept to a functional component of real-world materials discovery. While hardware limitations like embedding overhead, qubit connectivity, and finite coherence, still constrain problem size, the QUBO formulations themselves are hardware-portable and can directly map onto next-generation annealers or gate-based QAOA implementations. As these technologies mature, integrating physically informed descriptors with quantum-optimized models will enable increasingly autonomous, interpretable, and generalizable discovery pipelines. In this sense, the QaML framework provides not just a set of results, but a blueprint for merging quantum optimization with data-driven materials design.

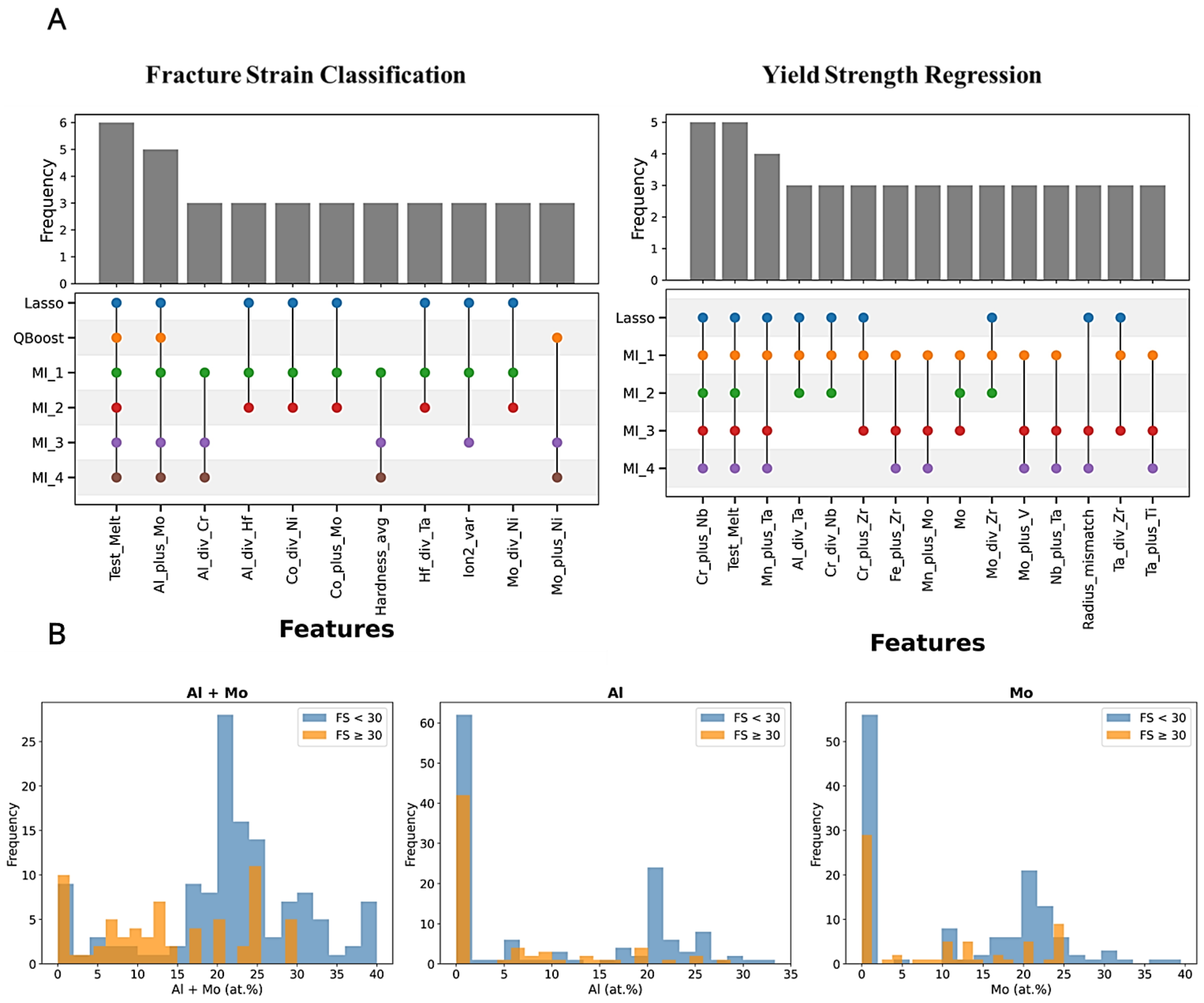

**Figure 6. Feature selection overlaps and feature–property relationships.** (a) UpSet-style plots comparing features selected by Lasso, QBoost, and Q-Mutual Information [MI_1 = ($\lambda_1 = 0.01$, $\lambda_2 = 0.01$), MI_2 = ($\lambda_1 = 0.01$, $\lambda_2 = 0.1$), MI_3 = ($\lambda_1 = 0.1$, $\lambda_2 = 0.01$), MI_4 = ($\lambda_1 = 0.1$, $\lambda_2 = 0.001$)] for classification (left) and regression (right). Only features chosen by at least two selectors are displayed. Bar heights indicate selection frequency, while dots below mark co-occurrence across methods. The limited overlap highlights that predictive performance depends on complementary feature combinations rather than a single shared subset, underscoring the need to evaluate complete feature sets. (b) Representative feature distributions for fracture-strain classification, illustrating the compositional dependence of ductility. These thresholds are dataset-specific.

# METHODS

## Experimental Methods

High- purity raw elements (>99.99 wt %) were arc melted in a water-cooled copper crucible using a 400-A current. Each melt lasted 30s, with at least five melts performed to ensure chemical homogeneity, and the ingots were flipped between each melt. The ingots were later polished for XRD. Parts of the ingots were also suction-cast into copper molds to produce a rod-shaped sample (3 mm diameter and 6 mm height) for compression testing.

The XRD measurements were conducted on the polished sample by a PANalytical Empyrean diffractometer with Cu Kα radiation and a scanning rate of approximately 0.15 degrees/s.

Following fabrication, the nominal composition and elemental homogeneity of CCA Al8Cr38Fe50MnTi2 were confirmed using energy-dispersive X-ray spectroscopy (EDS) area measurements and backscattered electron (BSE) imaging completed on the polished sample by the Phenom XLG2 scanning electron microscope (SEM). Per the machine's specifications, a "point" spot size was utilized for all measurements and images taken, alongside a working distance of 6 mm and an accelerating voltage of 15 kV. EDS area scans were repeated across different regions of the alloy's surface in order to determine global homogeneity.

Cylindrical samples with a diameter of 3 mm and a gauge length of 6 mm were used for compression testing. Compression tests were completed using uniaxial compression at room temperature using a computer-controlled 810 Material Test System (MTS) servo-hydraulic load frame with an initial strain rate of $1 \times 10^{-3} s^{-1}$. The load and displacement values of the testing system were recorded without specimens, and these displacement values were subtracted from the measured displacement values in compression tests at their corresponding load. This method excludes the elastic responses of the load frame, load cell, and sample grips. Then, the deformation of the specimens could be calculated.

To determine a measure of corrosion resistance in an engineering relevant environment, electrochemical characterization of CCA Al8Cr38Fe50MnTi2 and 304 grade stainless steel was performed in two solutions of increasing harshness. Namely, 0.1 M $H_2SO_4$ (pH ~ 0.76) with either 0.25 M NaCl or 0.5 M NaCl added were utilized to probe pitting resistance in low-pH conditions. 304 grade stainless steel was selected as a relevant and reasonable comparison to Al8Cr38Fe50MnTi2 due to its similarly high Cr content and lack of Mo often present in other stainless steel grades, which has long been known to increase pitting resistance[80]. To isolate pitting as the passive film breakdown mechanism, a scanning droplet cell (SDC, Biologic M470) was employed, which utilizes a freestanding solution droplet as a probe, avoiding the propensity of O-ring-based area normalization to cause confounding premature crevicing failure at the O-ring-sample interface[81]. Using the SDC, a 3-electrode cell was constructed, with the sample serving as working electrode (WE), a platinum wire serving counter electrode (CE), and a silver-silver chloride reference electrode (Ag-AgCl, RE). Solution was deaerated using $N_2$ gas prior to and during testing, and solution was consistently flowed through the droplet during all measurements. Finally, each sample was ground to a #1200 finish using SiC based polishing paper before rinsing under IPA to degrease and remove errant SiC particles from the alloy surface before every test.

To compare the corrosion resistance of Al8Cr38Fe50MnTi2 and 304 grade stainless steel, open-circuit potential (OCP) and potentiodynamic (PD) experiments were undertaken, both controlled and monitored using a Biologic SP-300 potentiostat. First, each alloy underwent 15 minutes of solution exposure at open circuit potential (OCP), where $E_{OCP}$ was measured over time, in order to allow each alloy's surface to become equilibrated with each solution. Following OCP, PD experiments were performed between -50 mV relative to alloy $E_{OCP}$ to either 1.35 $V_{SHE}$ (if the sample didn't pit), or a potential at which a current limit is triggered (i.e. if the sample prematurely failed by pitting). A scan rate of 1.667 mV/s was used for all PD experiments.

Finally, all resulting OCP and E-log($i$) curves underwent a boxcar (moving average) filter with a window size of 5 points before plotting in order to increase signal to noise ratio, as in similar works[82,83]

## Data Curation and Preprocessing

A comprehensive dataset was compiled from established high-entropy alloy databases[84–86] ,focusing on single-phase BCC HEAs. Compression data were prioritized due to their greater availability, as tensile measurements are less frequently reported for BCC systems.

The final dataset included 269 single-phase BCC for yield strength, and 153 BCC alloys for fracture strain. These entries encompassed diverse processing states (as-cast, annealed, and others) and test temperatures between 25 °C and 1000 °C. When multiple measurements were reported for a given alloy, their values were averaged to ensure consistency.

Compositions spanned a broad chemical space including Al, Nb, Ta, Ti, Zr, V, Mo, Cr, Fe, Mn, Ni, Co, Cu, and Hf, providing a representative basis for modeling mechanical behavior across BCC-structured HEAs.

Database can be found at Table S4 and Table S5.

## Quantum Annealing Framework

Quantum annealing was employed to solve the Quadratic Unconstrained Binary Optimization (QUBO) problems derived from feature selection, support-vector training, and neural-network pruning. The quantum processor evolves under a time-dependent Hamiltonian

$$H(t) = A(t)H_D + B(t)H_P, \qquad 5$$

where $H_D = -\sum_i \sigma_i^x$ is the transverse-field ("driver") Hamiltonian and $H_P$ encodes the optimization objective. The coefficients $A(t)$and $B(t)$ define the annealing schedule, decreasing and increasing respectively over the anneal duration. When the evolution proceeds adiabatically, the system remains in its instantaneous ground state and terminates in the ground state of $H_P$, corresponding to the optimal solution[4,5].

For binary optimization, $H_P$ is expressed in Ising form as

$$H_P = \sum_i h_i \sigma_i^z + \sum_{i<j} J_{ij} \sigma_i^z \sigma_j^z, \qquad 6$$

where $\sigma_i^z$ are Pauli-Z operators, $h_i$ represent local fields, and $J_{ij}$ denote pairwise couplings between qubits. The eigenvalues of $\sigma_i^z(\pm 1)$ encode binary decision variables $s_i$, which can be mapped to Boolean variables $x_i \in \{0,1\}$ through $s_i = 2x_i - 1$. This yields the QUBO formulation

$$E(x) = x^{\mathrm{T}} Q x, \qquad 7$$

With $Q$ containing both linear $(h_i)$ and quadratic $(J_{ij})$ coefficients. Solving the Ising Hamiltonian for its ground state is thus equivalent to minimizing the QUBO energy $E(x)$, which is directly implemented on the D-Wave Advantage quantum annealer.

## Quantum and Classical Annealing: Hardware and Implementation

Quantum annealing experiments were performed on the D-Wave Advantage system (Zephyr and Pegasus topologies) using the Ocean SDK. Minor embedding was carried out using D-Wave's default heuristic algorithm, with chain strengths scaled to the maximum logical coupling and verified through chain-break statistics. Each QUBO instance was solved with $n_{reads} = 1{,}000$, and anneal times $t_a = 100_{\mu s}$. Majority-vote chain repair was applied, and the lowest-energy feasible sample was retained for analysis.

In selected high-dimensional cases, particularly for the QSVM tasks, D-Wave's Hybrid Leap solver was used to mitigate embedding overhead while maintaining solution quality.

Classical Simulated Annealing (SA) baselines were executed using the `neal`[87] package from the Ocean SDK, with the same number of reads ($n_{reads} = 1{,}000$) and default temperature schedules. All SA computations were performed on the University of Virginia Rivanna High-Performance Computing cluster, equipped with Intel Xeon Platinum CPUs.

## QUBO-Based Feature Selection (QBoost and Q-MI)

Two quantum-assisted feature selection algorithms were implemented: QBoost and Quantum Mutual Information (Q-MI). Both methods cast feature selection as a discrete optimization problem formulated in QUBO form.

### QBoost

QBoost [59–62] is an ensemble-based selector in which each weak learner corresponds to a single feature descriptor. Depth-1 decision stumps $h_j: \mathbb{R} \to \{-1, +1\}$ were trained within each training fold to prevent data leakage, with one stump per descriptor $x^{(j)}$. A binary variable $\omega_i \in \{0,1\}$ denotes the inclusion of the corresponding feature. Given training pairs $\{(x_s, y_s)\}_{s=1}^{S}$ with $y_s \in \{0,1\}$, QBoost minimizes the sparse squared-loss objective:

$$\min_{\omega \in \{0,1\}^N} \left[ \sum_{s=1}^{S} \left( y_s - \sum_{i=1}^{N} \omega_i h_i(x_s) \right)^2 + \lambda \|\omega\|_0 \right] \quad 8$$

where $\lambda$ controls the sparsity of the selected subset. The optimal solution $\omega$defines a compact, interpretable feature mask $\{j: \omega_j = 1\}$ that is subsequently used as input to downstream classifiers. The regularization parameter $\lambda$was tuned via 5-fold cross-validation on the training data.

### Quantum Mutual Information (Q-MI)

Mutual information (MI) quantifies the dependence between random variables, capturing both linear and non-linear correlations. For discrete variables $X$ and $Y$

$$I(X;Y) = \sum_{y \in Y} \sum_{x \in X} p(x,y) \log \frac{p(x,y)}{p(x)p(y)}, \quad 9$$

where $p(x)$and $p(y)$ are marginal distributions and $p(x,y)$ is the joint distribution. In feature selection, $I(X_i;Y)$ represents the relevance of feature $X_i$ to the target variable, while $I(X_i;X_j)$ quantifies redundancy between feature pairs. The feature subset is obtained by solving the QUBO[63]:

$$E(\omega) = -\left[\sum_i \omega_i I(X_i;Y) - \lambda_1 \sum_{i<j} \omega_i \omega_j I(X_i;X_j) - \lambda_2 \sum_i \omega_i\right]. \quad 10$$

where the first term promotes features with strong target relevance, the second penalizes redundancy, and the third enforces compactness. The trade-off parameters $\lambda_1$and $\lambda_2$were optimized via 5-fold cross-validation. All MI scores were computed within folds to avoid information leakage.

For each candidate descriptor, feature–target relevance, $I(X_i;Y)$, was estimated using scikit-learn's `mutual_info_classif`[88] (classification) or `mutual_info_regression`[88] (regression) functions applied to standardized training data. The pairwise feature redundancies, $I(X_i;X_j)$, were calculated using `mutual_info_regression`, yielding a symmetric redundancy matrix used to construct the final QUBO.

## Recursive Batching strategy

Because the QUBO dimension in both QBoost and Q-MI scales with the number of candidate features $M$, a recursive batching strategy was implemented to maintain compatibility with the finite qubit connectivity of current quantum annealers (D-Wave Pegasus & Zephyr topologies). The full feature pool was divided into subsets of size $k$, and each sub-QUBO was solved independently using both quantum annealing (QA) on the D-Wave Advantage system and simulated annealing (SA) for classical benchmarking. Features selected in each batch were pooled, re-partitioned, and re-evaluated iteratively until convergence or until the remaining pool size fell below $k$. To avoid empty selections, the most relevant feature (highest relevance score) was retained whenever a batch produced no valid subset.

Batch sizes of $k = 20$, $50$, and $100$ were evaluated to assess scaling behavior and ensure consistency across problem sizes.

## Baseline Machine-Learning Models

Three baseline algorithms: Random Forests (RF), Support Vector Machines (SVM), and feed-forward Neural Networks (NN), were implemented for both regression and classification tasks using scikit-learn[88]. Each model was optimized via the five-fold cross-validated grid search, with the hyperparameter ranges automatically scaled according to dataset size (“small,” “medium,” or “large”) to balance model expressiveness and computational cost.

For Random Forests, the grid spanned the number of estimators ($n_{\text{estimators}} = 100 – 1000$), maximum tree depth ($10 – 40$ or unrestricted), and minimum samples per split ($2 – 10$).
For SVMs, both linear and radial basis function (RBF) kernels were tested, with penalty parameter $C \in [0.1,100]$, margin tolerance $\varepsilon \in [0.01,1.0]$ for regression, and kernel coefficient $\gamma \in \{\text{scale,auto}\}$.
For Neural Networks, multi-layer perceptron (MLPs) comprising two to four hidden layers with 32 – 512 neurons per layer were trained using $\alpha \in [10^{-4}, 10^{-1}]$, learning rates between $5 \times 10^{-4}$ and $3 \times 10^{-3}$, early stopping, and a maximum of 3,000 iterations.

All input features were standardized using StandardScaler. Model performance was evaluated on held-out test sets using the coefficient of determination ($R^2$) and root-mean-square error (RMSE) for regression, and accuracy and weighted F1-score for classification. The best-performing hyperparameters and their corresponding training and test metrics were recorded for downstream comparison.

The neural-network models served primarily as proxy architectures to identify suitable layer configurations and regularization parameters prior to full-scale training and quantum-annealing-based pruning in TensorFlow.

## Quantum Support Vector Machines (Classification and Regression)

### Classification

For support vector classification, the dual optimization problem was reformulated under the Karush–Kuhn–Tucker (KKT) conditions as

$$L = \frac{1}{2}\sum_{i,j} \alpha_i \alpha_j y_i y_j \phi(x_i, x_j) - \sum_i \alpha_i \qquad 11$$

subject to $0 \le \alpha_i \le C$ and $\sum_i \alpha_i y_i = 0$, where $\alpha_i$ are the dual coefficients, $y_i \in \{-1, +1\}$ the class labels, $\phi(\cdot,\cdot)$ the kernel function, and $C$ the box constraint parameter.

Each coefficient $\alpha_i$ was discretized into $K$ binary variables using a base-$B$ encoding,

$$\alpha_i = \sum_{k=0}^{K-1} B^k \, a_{iK+k}\,, \qquad a_{iK+k} \in \{0,1\}, \tag{12}$$

where each binary variable $a_{iK+k}$ represents a qubit. The equality constraint $\sum_i \alpha_i y_i = 0$ was incorporated as a quadratic penalty term with multiplier $\xi$, producing a fully unconstrained QUBO Hamiltonian solved via quantum annealing. Full derivations are provided in Willsch et al.[47]

Among the tested kernels, the radial basis function (RBF) kernel yielded the most stable and accurate performance.

**Regression**

For support vector regression, the dual problem was similarly expressed under the KKT conditions as

$$L = \frac{1}{2}\sum_{i,j}(\alpha_i - \alpha_i^*)(\alpha_j - \alpha_j^*)\phi(x_i, x_j) + \varepsilon\sum_i(\alpha_i - \alpha_i^*) \; - \sum_i y_i(\alpha_i - \alpha_i^*) \tag{13}$$

subject to $0 \leq \alpha_i, \alpha_i^* \leq C$ and $\sum_i(\alpha_i - \alpha_i^*) = 0$, where $\varepsilon$ defines the margin tolerance within which prediction errors are ignored.

Each paired Lagrange multiplier $(\alpha_i, \alpha_i^*)$ was discretized into $K$ binary coefficients using base-$B$ encoding:

$$\alpha_i = \sum_{k=0}^{K-1} B^k \, q_{iK+k}\,, \qquad \alpha_i^* = \sum_{k=0}^{K-1} B^k \, q_{NK+iK+k}\,, \qquad q_j \in \{0,1\}. \tag{14}$$

The equality constraint $\sum_i(\alpha_i - \alpha_i^*) = 0$ was incorporated as a quadratic penalty term, yielding a fully unconstrained QUBO Hamiltonian suitable for quantum annealing.

This formulation follows the same general structure recently reported by Dalal et al.[48] who independently developed a similar QUBO representation for support vector regression. The complete derivation and implementation details are nevertheless reproduced here for clarity and reproducibility (see Supplementary Information Section E and Section D).

For regression, the linear kernel provided the best predictive performance and computational stability.

### QUBO-Based Neural-Network Pruning

Full mathematical derivations of the pruning objective, QUBO formulation, and solver implementation (Fig. S2) (including parameter tuning, annealing configuration, and retraining protocol) are detailed in the Supplementary Information.

#### FUNDING

The present work is supported by the Office of Naval Research under Grant No. N00014-23-1-2441. NG and PKL acknowledge support from the National Science Foundation under Grant No. DMR-2226508.

#### AUTHOR CONTRIBUTIONS

Conceptualization: D.I.H., I.K., G.-W.C., J.P. Methodology: D.I.H. Software & Computational Implementation: D.I.H., H.J. Investigation (Mechanical & Microstructural Experiments): N.G., D.I.H. Investigation (Corrosion Testing & Electrochemistry): P.C. Writing – Original Draft: D.I.H. Writing – Review & Editing: D.I.H.,J.P., P.K.L., J.R.S., P.C., G.-W.C., I.K. Supervision: I.K., G.-W.C., P.K.L., J.R.S., J.P. Funding Acquisition: P.K.L., J.R.S., J.P.

#### COMPETING INTERESTS

The authors declare no competing interests.

#### MATERIALS & CORRESPONDENCE

Correspondence and requests for materials should be addressed to Diego Ibarra Hoyos (di8pd@virginia.edu) or Joseph Poon (sjp9x@virginia.edu).

# Supplementary Information:
# From Quantum Annealing to Alloy Discovery: Towards Accelerated Design of High-Entropy Alloys.

## Section A: Features

These are the total features used in this study (see Ibarra-Hoyos, et al[36]. for detailed definitions and calculation methods)

1. Thermodynamic Properties

- Mixing Enthalpy (Hmix): The enthalpy changes when constituent elements mix; relates to alloy- formation energy and phase stability.
- Mixing Entropy (Mix_Entropy): The configurational entropy due to multiple elements; stabilizes solid-solution phases.
- Strain Energy (Strain_Energy): Stored elastic energy in the lattice, often related to distortion and strengthening.
- Residual Strain (Residual_Strain): Remaining internal strain after processing or solidification.

2. Electronic Properties

- Average Valence Electron Concentration (VEC_avg): The mean number of valence electrons per atom; determines phase stability (e.g., FCC or BCC structure).
- Variance in Valence Electron Concentration (VEC_var): Variability of VEC among elements; indicates electronic heterogeneity.
- Average First Ionization Energy (Ion1_avg): Energy to remove the first electron; relates to bonding and chemical stability.
- Variance in First Ionization Energy (Ion1_var): Spread of first ionization energies among elements.
- Average Second Ionization Energy (Ion2_avg): Energy to remove the second electron; related to cation formation.
- Variance in Second Ionization Energy (Ion2_var): Variation in second ionization energy across alloy components.
- Average Third Ionization Energy (Ion3_avg): Energy to remove the third electron; linked to deeper electronic bonding strength.
- Variance in Third Ionization Energy (Ion3_var): Spread of third ionization energies

among constituent elements.

- Electronegativity Mismatch (Electronegativity_mismatch): Difference in electronegativity between elements; influences bonding, charge transfer, and lattice distortion.

3. Structural and Atomic Geometry Properties

- Average Atomic Density (Atomic_Density_avg): Mean number of atoms per unit volume; connected to packing density and lattice parameter.
- Variance in Atomic Density (Atomic_Density_var): Variation in atomic density among elements; indicates distortion.
- Atomic Radius Mismatch (Radius_mismatch): Relative difference in atomic radii; affects solid-solution strengthening.
- Atomic Size Misfit (Atomic_size_misfit): Quantitative measure of atomic size differences; affects lattice strain.
- Mean Square Misfit (Mean_square_misfit): Squared measures of atomic size differences averaged over elements.

4. Mechanical and Elastic Properties

- Hill Average Shear Modulus (Shear_hill): Effective shear modulus computed using Hill's average (between Voigt and Reuss bounds).
- Hill Average Young's Modulus (Youngs_hill): Effective Young's modulus estimated by Hill's method.
- Hill Average Bulk Modulus (Bulk_hill): Effective bulk modulus (resistance to compression) based on Hill's averaging.
- Poisson's Ratio (Poisson): Ratio of lateral to axial strain; indicator of ductility and elasticity balance.
- Average Hardness (Hardness_avg): Mean hardness value from mechanical testing (e.g., Vickers or Rockwell).
- Hardness Variance (Hardness_var): Variability in hardness measurements; reflects microstructural uniformity.

5. Defect and Fault-Related Properties

- Weighted Stacking Fault Energy (SFE_weighted): Composition-weighted average stacking fault energy; affects dislocation behavior and deformation.
- Weighted Unstable Stacking Fault Energy (USF_weighted): Weighted average of

unstable stacking fault energy; relates to slip and twinning tendencies.

- Weighted Diffusivity (D_weighted): Composition-weighted atomic diffusion coefficient; influences microstructural evolution and creep resistance.

6. Experimental or Categorical Identifiers

- Ratio of Testing Temperature and Melting Temperature (Test_Melt): Dimensionless ratio ($\frac{T_{test}}{T_{melt}}$) indicating the homologous temperature during testing.

7. Compositional Features

- Elemental Composition and Ratios: Atomic fractions of each constituent element (e.g., Al, Cr, Nb) along with all unique pairwise ratios (e.g., Al/Cr, Al/Nb, Cr/Nb, …) and sums (e.g., Al+Cr, Cr+Nb, …), representing relative and combined elemental proportions in the alloy.

## Section B: Features Selected.

**Table S1 | Final feature subsets selected for best-performing models.** Feature subsets identified by each selection method and used in the top-performing models for (i) fracture strain classification and (ii) yield strength regression. For fracture strain classification, the QBoost-selected features yielded the highest accuracy when coupled with the Random Forest classifier. For yield strength regression, Q-MI–derived subsets achieved optimal performance after λ-tuning, while the Linear SVM performed best with features selected via Lasso regularization. Only the final subsets corresponding to the best models are listed here.

| Model | Property | Features |
|---|---|---|
| QBoost | Fracture Strain | $\frac{T_{test}}{T_{melt}}$, Poisson, Zr, Al+Mo, Al+V, Cr+Mo, Cu+Zr, Hf+Ti, Hf+Zr, Mn+Zr, Mo+Ni, Nb+Zr, Ta+Zr |
| Q-MI ($\lambda_1 = 1,\ \lambda_2 = 0.01$) | Yield Strength | $\frac{T_{test}}{T_{melt}}$, Ion2_avg, Radius_mismatch, Atomic_size_misfit, Cr+Nb, Fe+Zr, Mn+Mo, Mn+Ta, Mo+V, Nb+Ta, Ta+Ti |
| LASSO | Yield Strength | Hmix, $\frac{T_{test}}{T_{melt}}$, Ion2_var, Ion3_var, Electronegativity_mismatch, Radius_mismatch, Mix_Entropy, Al/Co, Al/Cu, Al/Hf, Al/Ta, Al/Zr, Co/Cr, Co/Mn, Co/Ti, Co/V, Cr/Mo, Cr/Nb, Cr/V, Fe/Nb, Fe/Ti, Hf/Mo, Hf/Ta, Hf/Ti, Hf/Zr, Mn/Ti, Mo/Ni, Mo/Ta, Mo/V, Mo/Zr, Nb/Ti, Nb/Zr, Ni/V, Ta/Ti, Ta/V, Ta/Zr, V/W, V/Zr, Al+Ta, Cr+Nb, Cr+Zr, Hf+Ta, Mn+Ta, Mo+Zr, Ni+Ta, Ni+Ti, W+Zr |
| Q-MI ($\lambda_1 = 0.1,\ \lambda_2 = 0.001$) | Yield Strength | $\frac{T_{test}}{T_{melt}}$, Hardness_var, VEC_avg, Atomic_Density_avg, Ion1_var, Ion2_avg, Ion2_var, Ion3_var, Shear_hill, Youngs_hill, Poisson, Electronegativity_mismatch, Radius_mismatch, Residual_Strain, Mix_Entropy, Atomic_size_misfit, Mean_square_misfit, Mo, Nb, Ta, Nb/Ta, Ta/Zr, Al+Nb, Al+Ti, Co+Mo, Co+Ta, Cr+Nb, Cr+Ta, Cr+Ti, Cr+Zr, Cu+Mo, Cu+Ta, Fe+Nb, Fe+Ta, |

| | |
|---|---|
| | Fe+Zr, Hf+Nb, Mn+Mo, Mn+Ta, Mo+Ta, Mo+Ti, Mo+V, Mo+W, Mo+Zr, Nb+Ni, Nb+Ta, Nb+W, Nb+Zr, Ni+Ta, Ni+Zr, Ta+Ti, Ta+V, Ta+W, V+W |

## Section C: Machine Learning Model Implementation Details

The Random Forest (RF), Support Vector Machine (SVM), and feedforward Neural Network (NN) models were implemented in scikit-learn. Hyperparameter optimization was performed using exhaustive grid search with five-fold cross-validation, where the search range was adaptively scaled according to dataset size. The following grids were used:

- **Small datasets** ($n < 200$):

RF → n_estimators ∈ $\{100, 300\}$, max_depth ∈ $\{None, 10, 20\}$, min_samples_split ∈ $\{2, 5\}$

SVM → kernel ∈ {linear, rbf}, C ∈ {0.1, 1, 3, 10}, ε ∈ {0.01, 0.1, 0.5}, γ ∈ {‘scale’, ‘auto’}

NN → hidden_layer_sizes ∈ $\{(64, 32), (64, 32, 16), (64, 64, 32), (128, 64, 32), (32, 32, 16), (64, 32, 16, 8)\}$, $\alpha \in \{10^{-4}, 10^{-3}, 10^{-2}, 10^{-1}\}$, learning_rate_init ∈ $\{5 \times 10^{-4}, 10^{-3}, 3 \times 10^{-3}\}$, tol ∈ $\{10^{-3}, 10^{-4}\}$, n_iter_no_change ∈ $\{10, 20, 30\}$.

- **Medium datasets** ($200 \leq n < 1000$):

RF → n_estimators ∈ $\{200, 500, 800\}$, max_depth ∈ $\{None, 10, 30\}$, min_samples_split ∈ $\{2, 5,10\}$

SVM → kernel ∈ {linear, rbf}, C ∈ {0.1, 1, 3, 10,30}, ε ∈ {0.01, 0.1, 0.5,1.0}, γ ∈ {‘scale’, ‘auto’}

NN → hidden_layer_sizes ∈ $\{(128, 64), (128, 64, 32), (128, 128, 64), (256, 128, 64), (64, 64, 32), (128, 64, 32, 16), (256, 128, 64, 32)\}$, $\alpha \in \{10^{-4}, 10^{-3}, 10^{-2}, \}$, learning_rate_init ∈ $\{10^{-3}, 3 \times 10^{-3}\}$, tol ∈ $\{10^{-3}, 10^{-4}\}$, n_iter_no_change ∈ $\{10, 20, 30\}$.

- **Large datasets** ($n \geq 1000$):

RF → n_estimators ∈ $\{300, 600, 1000\}$, max_depth ∈ $\{None, 20, 40\}$, min_samples_split ∈ $\{2, 5,10\}$

SVM → kernel ∈ {linear, rbf}, C ∈ { 1, 3, 10,30,100}, ε ∈ {0.01, 0.1, 0.5,1.0}, γ ∈ {‘scale’, ‘auto’}

NN → hidden_layer_sizes ∈∈ $\{(256, 128), (256, 128, 64), (256, 256, 128), (512, 256, 128), (128, 128, 64), (256, 128, 64, 32), (512, 256, 128, 64$
$\alpha \in \{10^{-4}, 10^{-3}, \}$, learning_rate_init ∈ $\{10^{-3}, 3 \times 10^{-3}\}$, tol ∈ $\{10^{-3}, 10^{-4}\}$, n_iter_no_change ∈ $\{10, 20, 30\}$.

All MLPs used max_iter = 3000 and early_stopping = True. Input features were standardized via StandardScaler before training.

Regression models were scored using $R^2$ and RMSE, and classification models via accuracy and weighted F1-score. The best hyperparameter combination and corresponding train/test metrics were recorded for all models. The NN models here were not used for final deployment but rather to identify the optimal network depth, width, and regularization scales prior to full TensorFlow retraining and QUBO-based pruning analysis.

**Table S2 | Mutual Information parameters.** Q-MI configurations that achieved the minimum QUBO energy across tested hyperparameter values. We examined batch sizes of $K = 20, 50,$and $100$for both simulated annealing (SA) and quantum annealing (QA), while the complete problem set was also evaluated using SA.

**Fracture Strain**

| $\lambda_1$ | $\lambda_2$ | Q-MI Model |
|---|---|---|
| 0.01 | 0.01 | QA K=50 |
| 0.01 | 0.1 | QA K =50 |
| 0.1 | 0.001 | QA K=50 |
| 0.1 | 0.01 | QA K=50 |

**Yield Strength**

| $\lambda_1$ | $\lambda_2$ | Q-MI Model |
|---|---|---|
| 0.01 | 0.01 | SA K=20 |
| 0.01 | 0.1 | SA K =20 |
| 0.1 | 0.001 | QA K=100 |
| 0.1 | 0.01 | QA K=20 |

## Section D: Quantum-Support Vector Machine-Regression Formulation

**QUBO Formulation.** The dual problem of the SVM regression, constructed to satisfy the KKT conditions, is given by

$$L = \frac{1}{2}\sum_{i,j}(\alpha_i - \alpha_i^*)(\alpha_j - \alpha_j^*)\phi(x_i, x_j) + \varepsilon\sum_{i}(\alpha_i - \alpha_i^*) - \sum_{i} y_i(\alpha_i - \alpha_i^*) \quad 15$$

$$\text{s.t.} \quad \sum_i(\alpha_i - \alpha_i^*) = 0 \text{ and } 0 \le \alpha_i, \alpha_i^* \le C$$

**Binary Encoding.** Each continuous dual variable, $\alpha_i$ and $\alpha_i^*$, is discretized into $K$ binary variables using base-$B$ encoding:

$$\alpha_n = \sum_{k=0}^{K-1} B^k \cdot q_{nK+k}\,, \qquad \alpha_n^* = \sum_{k=0}^{K-1} B^k \cdot q_{NK+nK+k}, \quad 16$$

where

- $n \in \{0,1,\ldots,N-1\}$: index over training examples
- $k \in \{0,1,\ldots,K-1\}$: index over bits per variable
- $q_i \in \{0,1\}$: binary decision variable (qubit)
- $B$: base of encoding (typically 2)
- $N$: number of training examples

The term $nK + k$ identifies the binary variable for $\alpha_n$, while $NK + nK + k$ accesses the bits of $\alpha_n^*$

**Substitution into the Dual Problem**. After encoding, the difference between dual variables becomes

$$\alpha_n - \alpha_n^* = \sum_{k=0}^{K-1} B^k(q_{nK+k} - q_{NK+nK+k}). \quad 17$$

Each component of the dual objective can then be expanded in terms of binary variables:

1. Kernel term (quadratic):

$$\frac{1}{2}\sum_{n,m}\left[\left(\sum_{k=0}^{K-1} B^k(q_{nK+k} - q_{NK+nK+k})\right)\left(\sum_{j=0}^{K-1} B^j(q_{nK+k} - q_{NK+nK+k})\right) \cdot \phi(x_n, x_m)\right]. \quad 18$$

2. Equality constraint penalty (quadratic):

$$\xi\left(\sum_n \delta_n\right)^2 = \xi\left(\sum_n \sum_{k=0}^{K-1} B^k(q_{nK+k} - q_{NK+nK+k})\right)^2 \quad 19$$

3. $\varepsilon$-insensitive tube (linear):

$$\sum_i (\alpha_i - \alpha_i^*) = \varepsilon \sum_n \sum_{k=0}^{K-1} B^k(q_{nK+k} + q_{NK+nK+k}) \quad 20$$

4. Target term (linear):

$$-\sum_i y_i(\alpha_i - \alpha_i^*) = -\sum_n y_n \sum_{k=0}^{K-1} B^k(q_{nK+k} - q_{NK+nK+k}) \quad 21$$

Since $q_i^2 = q_i$, all terms can be directly expressed in quadratic form.

**QUBO Matrix Construction.** Using composite indices, $(n, k, s)$, where $s = 0$ for $\alpha$ and $s = 1$ for $\alpha^*$, the QUBO matrix elements are:

Off-diagonal terms:

$$Q_{(n,k,s),(m,j,t)} = \begin{cases} \frac{1}{2}B^{k+j}\phi(x_n, x_m) + \xi B^{k+j}, & s = t \\ -\frac{1}{2}B^{k+j}\phi(x_n, x_m) - \xi B^{k+j}, & s \neq t \end{cases} \quad 22$$

Diagonal terms:

$$Q_{(n,k,s),(n,k,s)} = \frac{1}{2}B^{2k}\phi(x_n, x_n) + \frac{1}{2}\xi B^{2k} + \begin{cases} (\varepsilon - y_n)B^k, & s = 0 \\ (\varepsilon + y_n)B^k, & s = 1 \end{cases} \quad 23$$

The resulting QUBO can then be directly mapped to a quantum annealer for optimization.

## Section E: Quantum-Support Vector Machine-Regression Implementation

We adopted the *combined-regressor* approach analogous to the ensemble method of Willsch *et al.*[47], in which multiple regressors trained on disjoint subsets of the data are averaged to form a composite predictor.

Two levels of averaging were performed:

1. Intra-split averaging: Within each data split, the D-Wave solver returns multiple near-optimal solutions $\{\alpha^{(l,i)}, b^{(l,i)}\}$ corresponding to the lowest-energy configurations. These were averaged as

$$\bar{\alpha}_n^{(l)} = \frac{1}{M}\sum_{i=1}^{M} \alpha_n^{(l,i)}, \bar{b}^{(l)} = \frac{1}{M}\sum_{i=1}^{M} b^{(l,i)}, \qquad 24$$

where $M$ is the number of accepted low-energy samples.

2. Inter-split averaging: To incorporate all available training data, the dataset was partitioned into $L$ disjoint subsets. The coefficients and biases from each subset were averaged as

$$\alpha_n = \frac{1}{L}\sum_{l=1}^{L} \bar{\alpha}_n^{(l)}, b = \frac{1}{L}\sum_{l=1}^{L} \bar{b}^{(l)}. \qquad 25$$

The final combined regressor is therefore expressed as

$$F(x) = \frac{1}{L}\sum_{l=1}^{L}\left[\sum_{n}\left(\bar{\alpha}_n^{(l)} - \bar{\alpha}_n^{*(l)}\right) K\left(x_n^{(l)}, x\right) + \bar{b}^{(l)}\right], \qquad 26$$

where $K(\cdot,\cdot)$ is the kernel function used in the QUBO formulation and $\bar{\alpha}_n^{*(l)}$ denotes the complementary dual variables for the regression task.

However, this approach ultimately produced $R^2$ values in the range of 0 – 0.1, indicating that no meaningful regression trend was captured. The poor performance arises from the inherent doubling of binary variables required to encode both $\alpha$ and $\alpha^*$in the regression formulation. This expansion significantly increases the number of qubits and, consequently, reduces the number of data points that can be embedded per batch. The resulting small effective training sets prevent the model from learning any statistical correlation, rendering the quantum regression approach currently unfeasible with existing hardware constraints.

## Section F: Neural-Network-Pruning Formulation

**Neural Network Architecture.** Each neuron within a feedforward neural network receives as input a set of activations $x_i$ originating from the preceding layer. For the first layer, $\mathbf{x}$ corresponds to the raw input features, while for subsequent layers $\mathbf{x}$ represent the activation function by the previous layer. Each neuron computes a weighted linear combination of these inputs, $z = \mathbf{w}^{\mathrm{T}}\mathbf{x} + b,$ where $\mathbf{w}$ denotes the vector of synaptic weights and $b$ the corresponding bias term. The scalar pre-activation value $z$ is subsequently transformed by a nonlinear activation function, $\sigma(z),$ to produce the neuron's output, $a = \sigma(z)$. In the present work, rectified linear units (ReLU) were used for all hidden layers, while the output layer employed a sigmoid activation for classification task and a linear activation for regression.

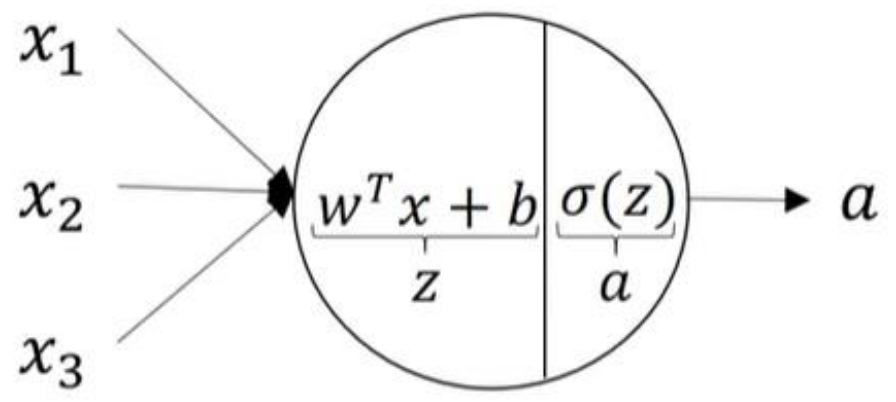

**Figure S1. Single neuron schematic.**

The neural network thus maps an input vector, $\mathbf{x}^{[l]}$, from layer, $l$, to an output activation vector, $\boldsymbol{a}^{[l+1]}$, according to

$$\boldsymbol{a}^{[l+1]} = \sigma\left(\mathbf{W}^{[l+1]}\boldsymbol{x}^{[l]} + \boldsymbol{b}^{[l+1]}\right), \quad 27$$

where $\mathbf{W}^{[l+1]} \in \mathbb{R}^{n_{l+1} \times n_l}$ and $\boldsymbol{b}^{[l+1]} \in \mathbb{R}^{n_{l+1}}$ denote the weight matrix and bias vector connecting layers, $l$ and $l+1$, respectively.

For a dataset containing $m$ training samples, the quantities introduced above can be extended to matrix form for compact representation of layer-wise operation. The activation matrix of layer, $l,$ is

$$\mathbf{A}^{[l]} = [\boldsymbol{a}^{[l](1)} \quad \boldsymbol{a}^{[l](2)} \quad \cdots \quad \boldsymbol{a}^{[l](m)}] \in \mathbb{R}^{n_l \times m}, \quad 28$$

where is the column, $\boldsymbol{a}^{[l](i)}$, represent the activation corresponding to the $i^{th}$-training example. The corresponding pre-activation matrix $\mathbf{Z}^{[l]}$ at layer $l$ is then defined as

$$\mathbf{Z}^{[l]} = \mathbf{W}^{[l]}\mathbf{A}^{[l-1]} + b^{[l]}\mathbf{1}^T, \qquad \mathbf{A}^{[l]} = \sigma\left(\mathbf{Z}^{[l]}\right), \quad 29$$

where $\mathbf{1} \in \mathbb{R}^m$ is a column vector of ones used to broadcast the bias across all training samples.

**Pruning Objective.** The objective of neuron pruning is to remove neurons that contribute the least to the representational power of the network, thereby creating a compact architecture that preserves performance. This serves two main purposes: (i) reducing computational and memory cost for deployment on resource-constrained hardware, and (ii) improving generalization by mitigating overfitting.

We determine which neurons to prune based on the following principles:

1. **Preservation of the current layer's activations,** ensuring that the reduced representation $\mathbf{A}'^{[l]}(\boldsymbol{\beta})$ remains close to the original $\mathbf{A}^{[l]}$
2. **Preservation of the next layer's pre-activations,** ensuring that the pruning of layer $l$ does not significantly distort $\mathbf{Z}^{[l+1]}$.

Pruning layer, $l$, alters both $\mathbf{A}^{[l]}$ and the subsequent pre-activation matrix, $\mathbf{Z}^{[l+1]} = \mathbf{W}^{[l+1]}\mathbf{A}^{[l]} + b^{[l+1]}\mathbf{1}^T$. The pruning problem is thus formulated as a Quadratic Unconstrained Binary Optimization (QUBO) that minimizes deviation in both $\mathbf{A}^{[l]}$ and $\mathbf{Z}^{[l+1]}$ simultaneously, while enforcing a constraint on the number of neurons retained.

A binary vector, $\boldsymbol{\beta} = \left[\beta_1, \ldots, \beta_{n_l}\right]^{\mathrm{T}}$, with $\beta_i \in \{0,1\}$, encodes whether neuron, $i$, is kept ($\beta_i = 1$) or pruned ($\beta_i = 0$). The goal is to find the optimal combination of $\beta_i$ values that minimizes the total energy

$$E(\boldsymbol{\beta}) = E_{[l]}(\boldsymbol{\beta}) + \mu E_{[l+1]}(\boldsymbol{\beta}) + E_{\text{penalization}}(\boldsymbol{\beta}) \qquad 30$$

where $\mu$ is a balancing parameter between intra- and inter-layer reconstruction.

**(1) Preservation of the current layer's activations.**

Deviation within layer $l$ is quantified via the Frobenius norm,

$$E_{[l]}(\boldsymbol{\beta}) = \left\|\mathbf{A}^{[l]} - \mathbf{A}'^{[l]}(\boldsymbol{\beta})\right\|_F^{2}. \qquad 31$$

Since pruning sets the activations of removed neurons to zero, $\mathbf{A}'^{[l]}(\boldsymbol{\beta}) = \mathbf{A}^{[l]} \cdot \boldsymbol{\beta}$, where $\cdot$ denotes elementwise multiplication along the neuron dimension. Expanding the expression yields

$$E_{[l]}(\boldsymbol{\beta}) = \sum_{k=1}^{m}\sum_{i=1}^{n_l}\left(a_{k,i}^{[l]} - \beta_i a_{k,i}^{[l]}\right)^2 = \sum_{i=1}^{n_l} s_i(1 - \beta_i), \qquad 32$$

where $s_i = \sum_{k=1}^{m} \left(a_{i,k}^{[l]}\right)^2$ measures the total activation strength of neuron, $i$, across all samples. Since $\sum_{i=1}^{n_l} s_i$ is constant, it can be dropped without affecting the optimization landscape, giving

$$E_{[l]}(\boldsymbol{\beta}) = -\sum_{i=1}^{n_l} s_i \beta_i \qquad 33$$

**(2) Preservation of next layer pre-activations**

Deviation in the next layer's pre-activations is similarly computed as

$$E_{[l+1]}(\boldsymbol{\beta}) = \left\|\mathbf{Z}^{[l+1]} - \mathbf{Z}'^{[l+1]}(\boldsymbol{\beta})\right\|_F^2 = \left\|\mathbf{W}^{[l+1]}\mathbf{A}^{[l]} - \mathbf{W}^{[l+1]}\mathbf{A}'^{[l]}(\boldsymbol{\beta})\right\|_F^2. \qquad 34$$

The bias term of $\boldsymbol{b}^{[l+1]}$ cancels out, as it is unaffected by pruning. Expanding yields

$$E_{[l+1]}(\boldsymbol{\beta}) = \sum_{i=1}^{n_l}\sum_{j=1}^{n_l}(1-\beta_i)(1-\beta_j)M_{i,j}, \qquad 35$$

where

$$M_{ij} = \sum_{r=1}^{n_{l+1}}\sum_{k=1}^{m} w_{r,i}^{[l+1]} w_{r,j}^{[l+1]} a_{i,k}^{[l]} a_{j,k}^{[l]} \qquad 36$$

encodes the interaction between neurons $i$ and $j$ as transmitted to the next layer. Expanding and simplifying, dropping constants, gives

$$E_{[l+1]}(\boldsymbol{\beta}) = 2\sum_{i=1}^{n_l}\beta_i \sum_{j=1}^{n_l} M_{i,j} + \sum_{i=1}^{n_l}\sum_{j=1}^{n_l} M_{i,j}\beta_i\beta_j \qquad 37$$

**(3) Cardinality Constraint**

To prune exactly $k$ neurons, a quadratic penalty term is added:

$$E_{\text{penalization}}(\boldsymbol{\beta}) = \lambda\left(\sum_{i=1}^{m}\beta_i - k\right)^2 = \lambda(1-2k)\sum_{i=1}^{m}\beta_i + 2\lambda\sum_{i\leq j}\beta_i\beta_j \qquad 38$$

where the constant, $\lambda k^2$, is omitted since it does not affect the optimization landscape.

**(4) Final QUBO formulation**

Combining the three energy terms yields the final QUBO expression:

$$E_{\text{QUBO}}(\boldsymbol{\beta}) = \sum_{i \le j} J_{ij}\beta_i\beta_j + \sum_{i=1}^{n_l} h_i\beta_i, \qquad 39$$

where the coefficients are

$$\begin{aligned} J_{ij} &= 2(\mu M_{ij} + \lambda), \\ h_i &= -s_i \; + \; \mu\left(-2\sum_{j=1}^{n_l} M_{i,j} + M_{i,i}\right) \; + \; \lambda(1 - 2k). \end{aligned} \qquad 40$$

The QUBO energy landscape defined by $J_{ij}$ and $h_i$ encapsulates the trade-off between information preservation and model sparsity. Minimization of $E_{\text{QUBO}}(\boldsymbol{\beta})$ is performed using either classical simulated annealing or quantum annealing, as described in the main text.

## Section G: Neural Network Pruning Computational Implementation

All pruning experiments, both for fracture strain classification and yield strength regression, were carried out using a unified Python workflow implemented in TensorFlow 2.15, interfaced with D-Wave's hybrid quantum annealing platform (dwave-system) and the dimod modeling framework. The scripts provide full reproducibility under fixed random seeds and deterministic TensorFlow kernels (`TF_DETERMINISTIC_OPS=1`).

**Network Architecture and Training.** Both models employed custom `MaskedDense` layers, which extend the standard dense layer by including a binary, non-trainable mask, $\boldsymbol{\beta} \in \{0,1\}^{n_l}$, which enables explicit control of neuron activation during pruning.

- **Fracture strain classification model:** architecture (32,32,16,1) with ReLU activations in hidden layers and sigmoid output. Trained for 500 epochs (batch=32) using Adam optimizer ($lr = 10^{-3}$) and binary cross-entropy loss.
- **Yield strength regression model:** architecture (64,64,32,1) with ReLU hidden layers and a linear output neuron. Trained for 600 epochs (batch = 32) with Adam optimizer and mean-squared-error (MSE) loss.

**Layer-wise Pruning Protocol.** After baseline training, specific layers were selected for pruning. The procedure was identical for both tasks and executed following these steps:

**(1) Extraction of activations:**

Compute $\mathbf{A}^{[l-1]}, \mathbf{Z}^{[l]}$, and $\mathbf{A}^{[l]}$ for the entire training set.

**(2) QUBO construction:**

Using the formalism,

$$E_{\text{QUBO}}(\boldsymbol{\beta}) = \sum_{i \le j} J_{ij}\beta_i\beta_j + \sum_{i=1}^{n_l} h_i\beta_i$$

with $J_{ij}$ and $h_i$ defined from the preservation objectives.

**(3) Automatic hyperparameter tuning:**

The weighting coefficients, $\mu$ and $\lambda$, are automatically inferred from the activation energies $s_i$ and coupling matrix, $M_{ij}$, via median-based scaling. This process avoids manual hyperparameter sweeps while maintaining numerical balance between reconstruction and sparsity penalties

**(4) Normalization and conditioning:**

Each activation matrix is column-centered and $l_2$-normalized by default to stabilize the matrix used in the QUBO construction

**(5) Annealing-based minimization:**

The resulting QUBO is mapped to a `dimod.BinaryQuadraticModel` and solved by either:

- Quantum Annealing, on D-Wave's Zephyr topology with 1000 reads and 100 $\mu s$ anneal time per read; or
- Simulated Annealing ,using `neal.SimulatedAnnealingSampler()` as a classical baseline under identical read count.

**(6) Mask application:**

The optimal binary vector, $\boldsymbol{\beta}^*$**,** minimizing $E_{\text{QUBO}}$ is applied to the corresponding `MaskedDense` layer, permanently deactivating pruned neurons

**(7) Retraining:**

The pruned network is recompiled and fine-tuned for 5 epochs (batch = 16) to allow minor weight adaptation.

**Layer Selection and Validation Strategy.** To determine which layers and how many neurons to prune, multiple layer-neuron combinations were explored. For each candidate configuration, the model was re-trained for 5 epochs on the training and validation subsets only, and its loss function was monitored. The test set remained strictly unseen throughout this tuning process and was used only once for the final performance evaluation after selecting the optimal pruning configuration.

This procedure ensures that pruning decisions were based solely on training-validation behavior, preventing information leakage from the test data and yielding an unbiased measure of post-pruning generalization.

**Table S3 | QUBO parameters and resulting energies for pruning optimization.** Summary of hyperparameters ($\mu$, $\lambda$) and final QUBO energy values obtained during neuron pruning for fracture strain classification and yield strength regression. Both Simulated Annealing (SA) and Quantum Annealing (QA) were performed under identical hyperparameter settings. In both cases, SA achieved slightly lower final energy minima compared to QA.

**Fracture Strain**

| Solver | QUBO Energy | $\mu$ | $\lambda$ |
|---|---|---|---|
| SA | 13.91272 | 0.094901 | 0.376867 |
| QA | 18.0041 | 0.094901 | 0.376867 |

**Yield Strength**

| Solver | QUBO Energy | $\mu$ | $\lambda$ |
|---|---|---|---|
| SA | 31.38308 | 0.027573 | 0.228062 |
| QA | 34.87828 | 0.027573 | 0.228062 |

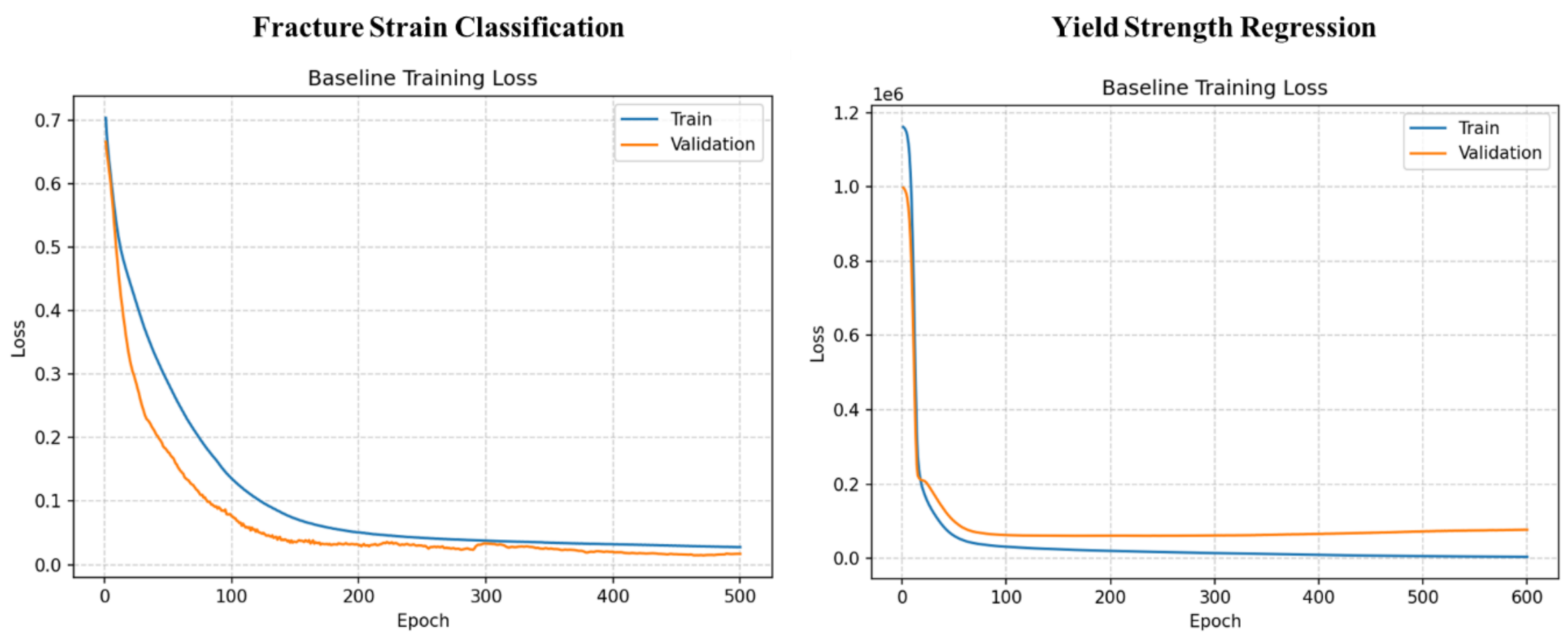

**Figure S2. Baseline neural-network training performance. Training** and validation loss curves for the unpruned models prior to QUBO-based pruning. *(Left)* Fracture strain classification network trained for 500 epochs using binary cross-entropy loss. *(Right)* Yield strength regression network trained for 600 epochs using mean-squared-error loss. Both models show stable convergence.

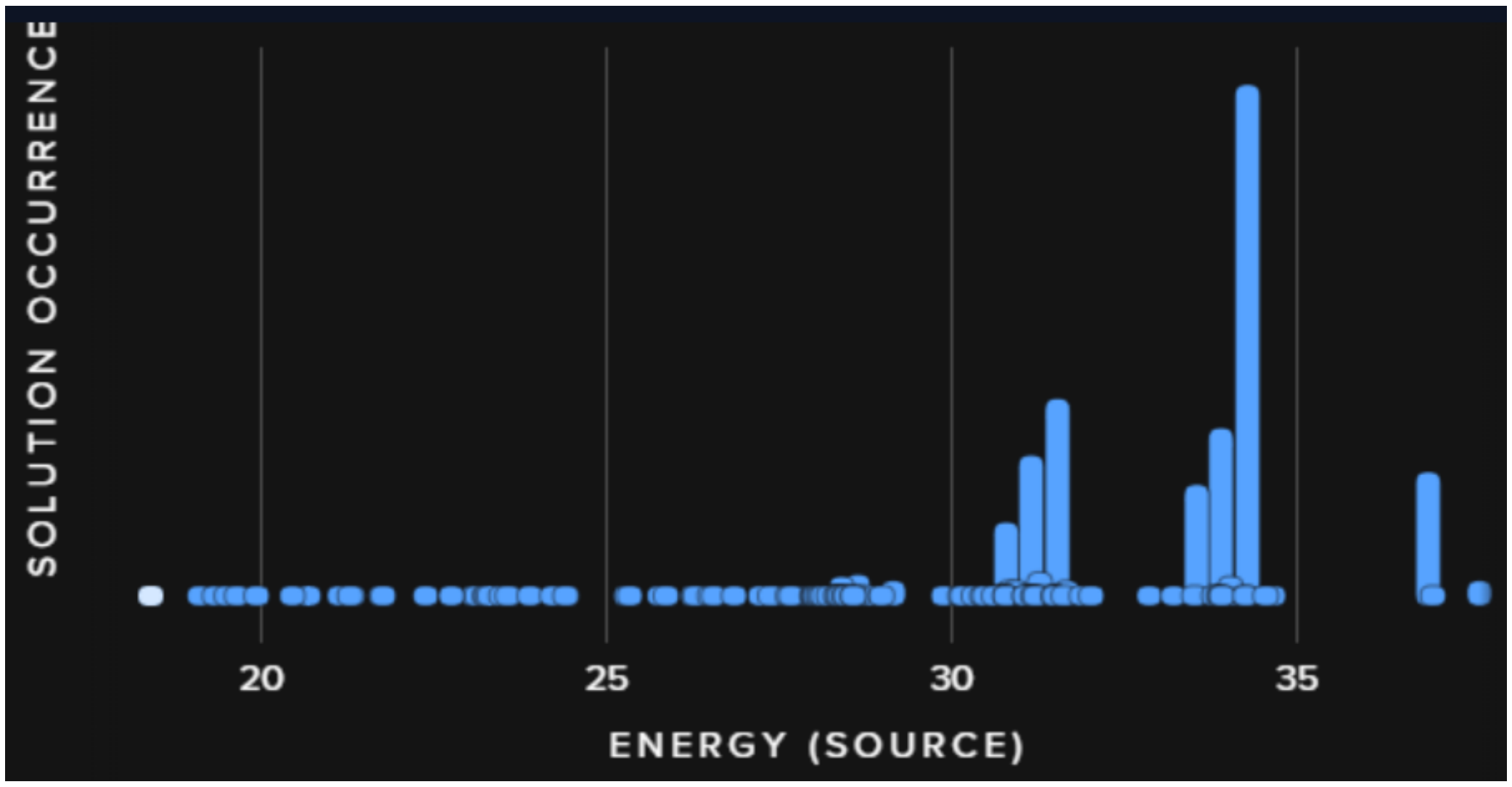

**Figure S3. Representative quantum annealing energy landscape**. Distribution of solution occurrences as a function of source energy for a representative pruning QUBO. Each point corresponds to a sampled solution from the quantum annealer, with lower energies indicating better objective minimization. The configuration with the minimum energy was selected as the optimal pruning mask.

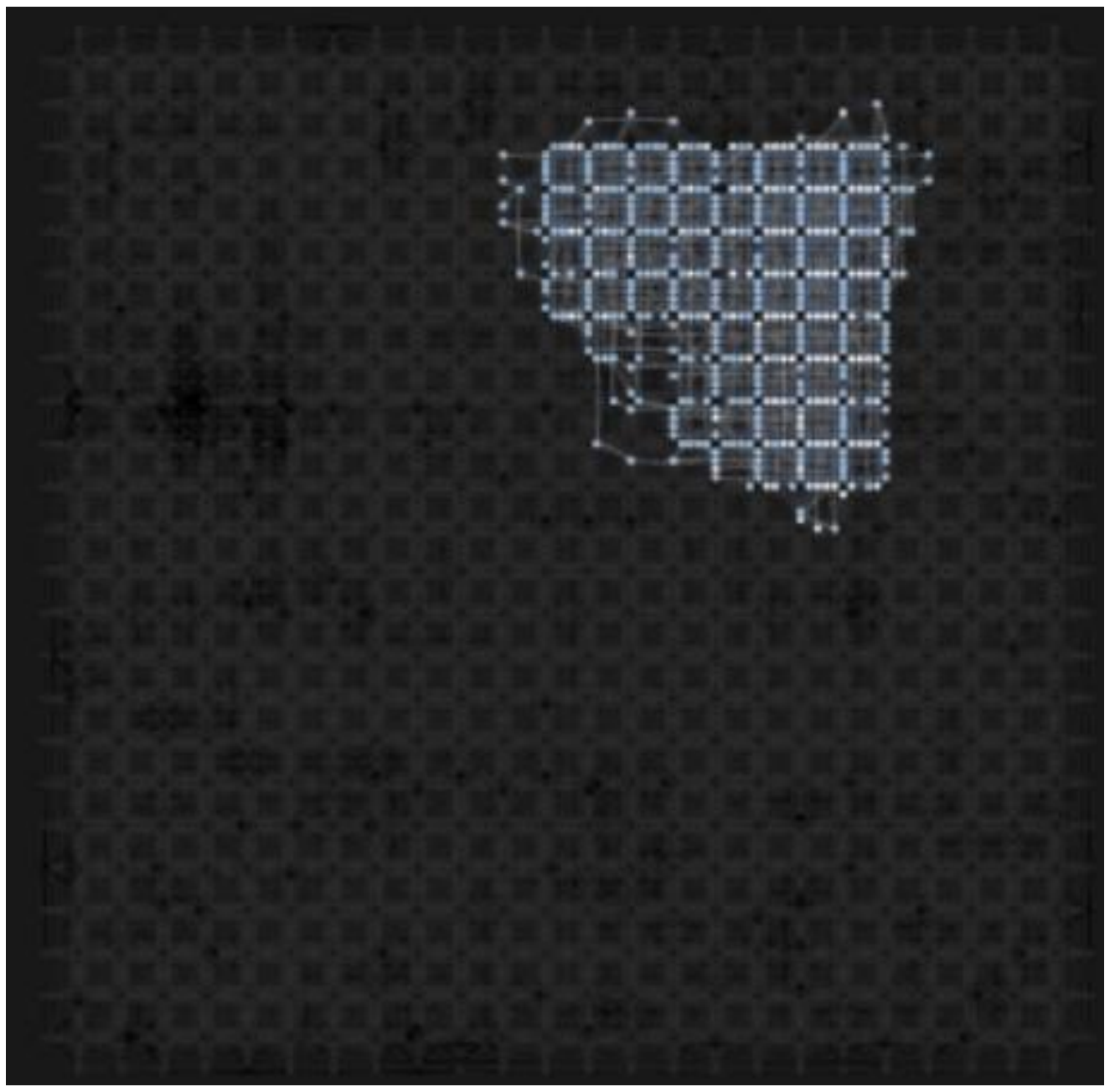

**Figure S4. Example of QUBO embedding on D-Wave quantum annealer**. Representative visualization of the pruning QUBO mapped onto the D-Wave Zephyr topology. Each node represents a physical qubit, and connected chains encode logical variables used during the annealing process.

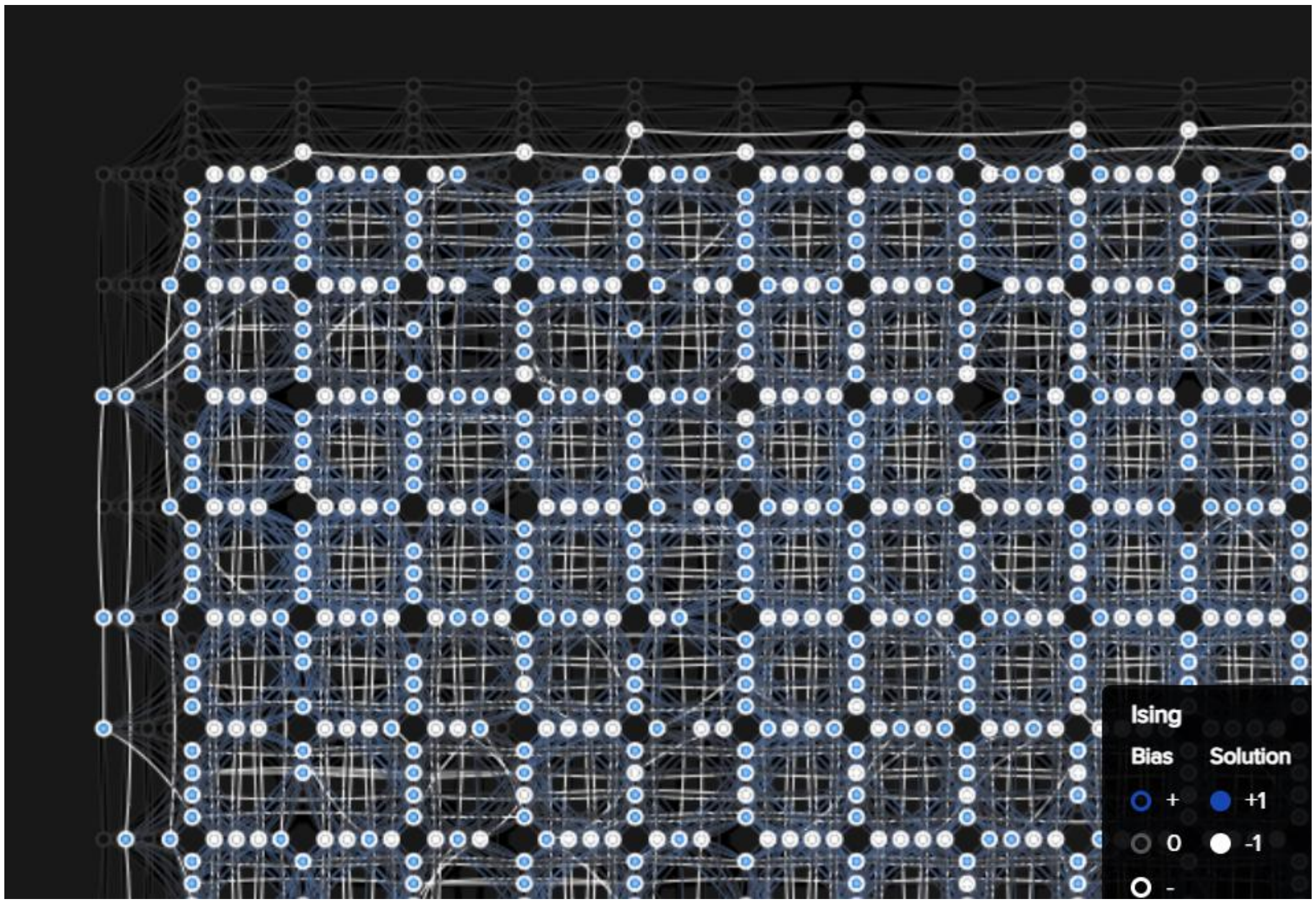

**Figure S5. Close-up view of QUBO embedding on D-Wave Zephyr topology.** Magnified visualization of the pruning problem embedded onto the D-Wave Advantage quantum annealer. The dense inter-qubit connectivity of the Zephyr graph enables mapping of higher-order interactions between logical variables during annealing.

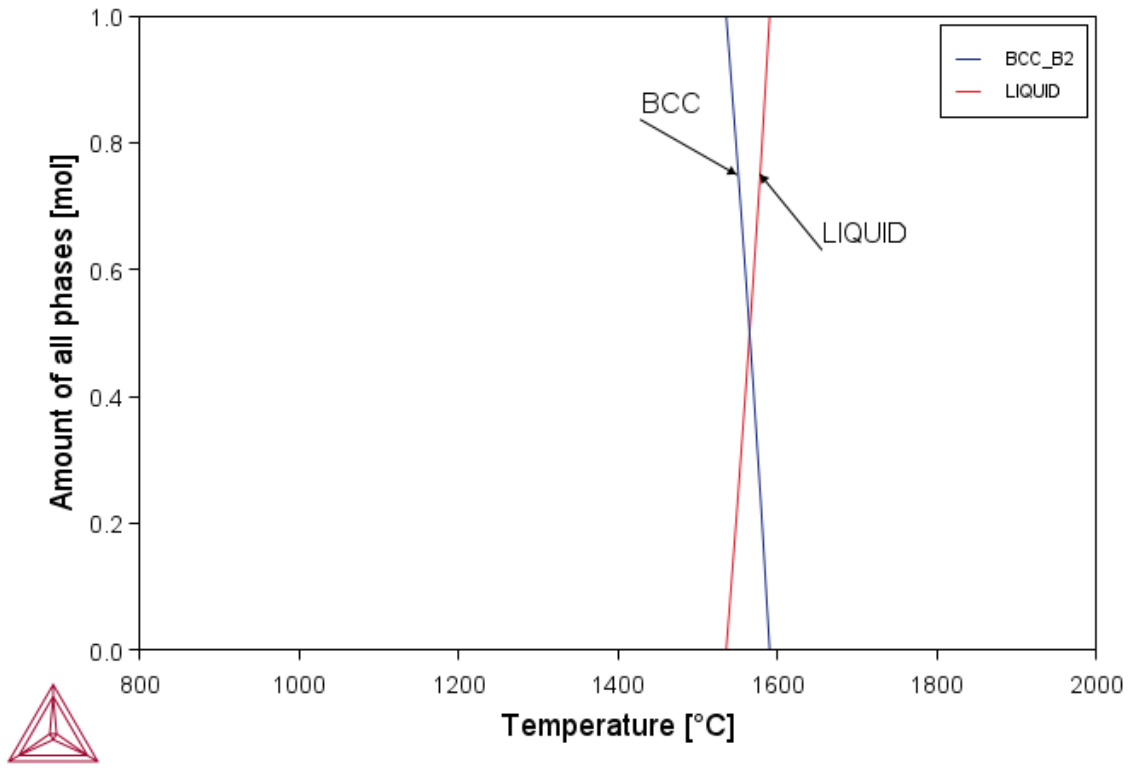

**Figure S6. CALPHAD of Al8Cr32Fe50Ti2Mn2 (at. %)**

**Table S4.** Dataset comprising Yield Strength values (MPa) for BCC HEAs. The dataset includes columns for alloy composition, phase constitution, mechanical testing method, processing condition, testing temperature (°C), and corresponding yield strength values. In this context, "A" denotes alloys that have undergone Annealing, "AC" represents those in the As Cast condition, and "OTHER" refers to all other processing conditions. The mechanical testing method is specified as "C" for Compression testing.

| BCC Yield Strength Dataset | | | | | |
|---|---|---|---|---|---|
| **Alloy** | **Phases** | **Mechanical Testing** | **Processing Condition** | **Test Temperature C** | **Yield Strength (MPa)** |
| Al0.214Nb0.714Ta0.571Ti1V0.143Zr0.929 | BCC | C | OTHER | 25.0 | 1965.0 |
| Al0.214Nb0.714Ta0.571Ti1V0.143Zr0.929 | BCC | C | OTHER | 800.0 | 678.0 |
| Al0.214Nb0.714Ta0.571Ti1V0.143Zr0.929 | BCC | C | OTHER | 1000.0 | 166.0 |
| Al0.214Nb0.714Ta0.714Ti1Zr0.929 | BCC | C | OTHER | 25.0 | 1965.0 |
| Al0.214Nb0.714Ta0.714Ti1Zr0.929 | BCC | C | OTHER | 800.0 | 362.0 |
| Al0.214Nb0.714Ta0.714Ti1Zr0.929 | BCC | C | OTHER | 1000.0 | 236.0 |
| Al0.25Co1Cr1Cu1Fe1Mn1Ni1Ti1V1 | BCC | C | AC | 25.0 | 1465.0 |
| Al0.25Cr0.5Nb0.5Ti1V0.25 | BCC | C | AC | 25.0 | 1240.0 |
| Al0.25Mo1Nb1Ti1V1 | BCC | C | AC | 25.0 | 1250.0 |
| Al0.25Nb1Ta1Ti1V1 | BCC | C | AC | 25.0 | 1330.0 |
| Al0.2Mo1Ta1Ti1V1 | BCC | C | AC | 25.0 | 1021.0 |
| Al0.333Nb0.667Ta0.533Ti1V0.133Zr0.667 | BCC | C | OTHER | 25.0 | 2035.0 |
| Al0.333Nb0.667Ta0.533Ti1V0.133Zr0.667 | BCC | C | OTHER | 800.0 | 796.0 |
| Al0.333Nb0.667Ta0.533Ti1V0.133Zr0.667 | BCC | C | OTHER | 1000.0 | 220.0 |
| Al0.3Hf1Nb1Ta1Ti1Zr1 | BCC | C | AC | 25.0 | 1188.0 |
| Al0.3Nb1Ta0.8Ti1.4V0.2Zr1.3 | BCC | C | OTHER | 25.0 | 1965.0 |
| Al0.3Nb1Ta0.8Ti1.4V0.2Zr1.3 | BCC | C | OTHER | 800.0 | 678.0 |
| Al0.3Nb1Ta0.8Ti1.4V0.2Zr1.3 | BCC | C | OTHER | 1000.0 | 166.0 |
| Al0.4Hf0.6Nb1Ta1Ti1Zr1 | BCC | C | OTHER | 25.0 | 1841.0 |
| Al0.4Hf0.6Nb1Ta1Ti1Zr1 | BCC | C | OTHER | 800.0 | 796.0 |
| Al0.4Hf0.6Nb1Ta1Ti1Zr1 | BCC | C | OTHER | 1000.0 | 298.0 |
| Al0.4Hf0.6Nb1Ta1Ti1Zr1 | BCC | C | OTHER | 1200.0 | 89.0 |
| Al0.5Cr1Nb1Ti2V0.5 | BCC | C | AC | 25.0 | 1240.0 |
| Al0.5Hf1Nb1Ta1Ti1Zr1 | BCC | C | AC | 25.0 | 1302.0 |
| Al0.5Mo1Nb1Ti1V1 | BCC | C | AC | 25.0 | 1625.0 |
| Al0.5Nb1Ta1Ti1V1 | BCC | C | AC | 25.0 | 1014.0 |
| Al0.667Nb1Ta0.333Ti1Zr0.333 | BCC | C | OTHER | 25.0 | 1280.0 |
| Al0.667Nb1Ta0.333Ti1Zr0.333 | BCC | C | OTHER | 800.0 | 728.0 |
| Al0.667Nb1Ta0.333Ti1Zr0.333 | BCC | C | OTHER | 1000.0 | 403.0 |
| Al0.6Mo1Ta1Ti1V1 | BCC | C | AC | 25.0 | 962.0 |
| Al0.75Hf1Nb1Ta1Ti1Zr1 | BCC | C | AC | 25.0 | 1415.0 |
| Al0.75Mo1Nb1Ti1V1 | BCC | C | AC | 25.0 | 1260.0 |
| Al1.5Mo1Nb1Ti1V1 | BCC | C | AC | 25.0 | 500.0 |
| Al1Co0.5Cr0.5Cu0.5Fe0.5Ni0.5 | BCC | C | AC | 25.0 | 1620.0 |
| Al1Co0.5Cr0.5Cu0.5Fe0.5Ni0.5 | BCC | C | AC | 500.0 | 1120.0 |
| Al1Co0.5Cr0.5Cu0.5Fe0.5Ni0.5 | BCC | C | AC | 600.0 | 805.0 |
| Al1Co0.5Cr0.5Cu0.5Fe0.5Ni0.5 | BCC | C | AC | 700.0 | 567.0 |
| Al1Co0.5Cr0.5Cu0.5Fe0.5Ni0.5 | BCC | C | AC | 800.0 | 302.0 |
| Al1Co0.5Cr0.5Cu0.5Fe0.5Ni0.5 | BCC | C | AC | 900.0 | 214.0 |
| Al1Co0.5Cr0.5Cu0.5Fe0.5Ni0.5 | BCC | C | AC | 1000.0 | 116.0 |
| Al1Co0.5Cr0.5Cu0.5Fe0.5Ni0.5 | BCC | C | AC | 1100.0 | 79.0 |
| Al1Co1Cr1Fe1Mo0.1Ni1 | BCC | C | AC | 25.0 | 1804.0 |
| Al1Co1Cr1Fe1Nb0.1Ni1 | BCC | C | AC | 25.0 | 1641.0 |
| Al1Co1Cr1Fe1Ni1 | BCC | C | AC | 25.0 | 1184.6 |
| Al1Co1Fe1Ni1 | BCC | C | AC | 25.0 | 964.0 |
| Al1Cr0.5Nb1Ti1V1 | BCC | C | A | 25.0 | 1300.0 |
| Al1Cr0.5Nb1Ti1V1 | BCC | C | A | 600.0 | 1005.0 |
| Al1Cr0.5Nb1Ti1V1 | BCC | C | A | 800.0 | 640.0 |
| Al1Cr0.5Nb1Ti1V1 | BCC | C | A | 1000.0 | 40.0 |
| Al1Cr1Mo1Nb1Ti1 | BCC | C | AC | 400.0 | 1080.0 |
| Al1Cr1Mo1Nb1Ti1 | BCC | C | AC | 600.0 | 1060.0 |
| Al1Cr1Mo1Nb1Ti1 | BCC | C | AC | 800.0 | 860.0 |
| Al1Cr1Mo1Nb1Ti1 | BCC | C | AC | 1000.0 | 594.0 |
| Al1Cr1Mo1Nb1Ti1 | BCC | C | AC | 1200.0 | 105.0 |
| Al1Cr1Mo1Ti1 | BCC | C | A | 25.0 | 1100.0 |
| Al1Cr1Mo1Ti1 | BCC | C | A | 400.0 | 1070.0 |
| Al1Cr1Mo1Ti1 | BCC | C | A | 600.0 | 1020.0 |
| Al1Cr1Mo1Ti1 | BCC | C | A | 800.0 | 875.0 |
| Al1Cr1Mo1Ti1 | BCC | C | A | 1000.0 | 375.0 |
| Al1Cr1Mo1Ti1 | BCC | C | A | 1200.0 | 100.0 |
| Al1Mo1Nb1Ti1 | BCC | C | A | 25.0 | 1100.0 |
| Al1Mo1Nb1Ti1 | BCC | C | A | 600.0 | 520.0 |
| Al1Mo1Nb1Ti1 | BCC | C | A | 800.0 | 500.0 |
| Al1Mo1Nb1Ti1 | BCC | C | A | 1000.0 | 540.0 |
| Al1Mo1Nb1Ti1 | BCC | C | A | 1200.0 | 200.0 |
| Al1Mo1Nb1Ti1V1 | BCC | C | AC | 25.0 | 1375.0 |
| Al1Nb1.5Ta0.5Ti1.5Zr0.5 | BCC | C | OTHER | 25.0 | 1280.0 |
| Al1Nb1.5Ta0.5Ti1.5Zr0.5 | BCC | C | OTHER | 800.0 | 728.0 |
| Al1Nb1.5Ta0.5Ti1.5Zr0.5 | BCC | C | OTHER | 1000.0 | 403.0 |
| Al1Nb1Ta1Ti1 | BCC | C | AC | 25.0 | 1152.0 |
| Al1Nb1Ta1Ti1 | BCC | C | AC | 100.0 | 740.0 |
| Al1Nb1Ta1Ti1 | BCC | C | AC | 200.0 | 740.0 |
| Al1Nb1Ta1Ti1V1 | BCC | C | AC | 25.0 | 993.0 |
| Al1Nb1Ti1V1 | BCC | C | A | 25.0 | 1010.0 |
| Al1Nb1Ti1V1 | BCC | C | A | 600.0 | 795.0 |
| Al1Nb1Ti1V1 | BCC | C | A | 800.0 | 622.5 |
| Al1Nb1Ti1V1 | BCC | C | A | 1000.0 | 134.0 |
| Cr1Mo1Nb1Ti1 | BCC | C | A | 25.0 | 1630.0 |
| Cr1Mo1Nb1Ti1 | BCC | C | A | 200.0 | 1268.0 |
| Cr1Mo1Nb1Ti1 | BCC | C | A | 400.0 | 1115.0 |
| Cr1Mo1Nb1Ti1 | BCC | C | A | 600.0 | 1062.0 |
| Cr1Mo1Nb1Ti1 | BCC | C | A | 800.0 | 1058.0 |
| Hf0.25Nb0.125Ti1V0.5Zr0.5 | BCC | C | AC | 25.0 | 1115.0 |
| Hf0.25Nb0.25Ti1V0.5Zr0.5 | BCC | C | AC | 25.0 | 1065.0 |
| Hf0.25Nb0.25Ti1V0.5Zr0.5 | BCC | C | AC | 600.0 | 718.0 |
| Hf0.25Nb0.25Ti1V0.5Zr0.5 | BCC | C | AC | 800.0 | 135.0 |
| Hf0.25Nb0.375Ti1V0.5Zr0.5 | BCC | C | AC | 25.0 | 1025.0 |
| Hf0.25Nb0.5Ti1V0.5Zr0.5 | BCC | C | AC | 25.0 | 980.0 |
| Hf0.25Nb0.5Ti1V0.5Zr0.5 | BCC | C | AC | 600.0 | 859.0 |
| Hf0.25Nb0.5Ti1V0.5Zr0.5 | BCC | C | AC | 800.0 | 195.0 |
| Hf0.25Ti1V0.5Zr0.5 | BCC | C | AC | 25.0 | 1160.0 |
| Hf0.25Ti1V0.5Zr0.5 | BCC | C | AC | 600.0 | 405.0 |
| Hf0.25Ti1V0.5Zr0.5 | BCC | C | AC | 800.0 | 85.0 |
| Hf0.26Nb1Ta1Ti0.578Zr0.416 | BCC | C | AC | 25.0 | 845.0 |
| Hf0.26Nb1Ta1Ti0.578Zr0.416 | BCC | C | AC | 60.0 | 795.0 |
| Hf0.26Nb1Ta1Ti0.578Zr0.416 | BCC | C | AC | 100.0 | 765.0 |
| Hf0.26Nb1Ta1Ti0.578Zr0.416 | BCC | C | AC | 200.0 | 650.0 |
| Hf0.26Nb1Ta1Ti0.578Zr0.416 | BCC | C | AC | 300.0 | 590.0 |
| Hf0.4Nb1.54Ta1.54Ti0.89Zr0.64 | BCC | C | AC | 20.0 | 822.0 |
| Hf0.4Nb1.54Ta1.54Ti0.89Zr0.64 | BCC | C | AC | 60.0 | 795.0 |
| Hf0.4Nb1.54Ta1.54Ti0.89Zr0.64 | BCC | C | AC | 100.0 | 765.0 |
| Hf0.4Nb1.54Ta1.54Ti0.89Zr0.64 | BCC | C | AC | 200.0 | 650.0 |
| Hf0.4Nb1.54Ta1.54Ti0.89Zr0.64 | BCC | C | AC | 300.0 | 590.0 |
| Hf0.5Mo0.5Nb1Ti1Zr1 | BCC | C | AC | 25.0 | 1176.0 |
| Hf0.5Nb0.5Ta0.5Ti1.5Zr1 | BCC | C | AC | 25.0 | 903.0 |
| Hf0.5Nb0.667Ta0.333Ti1Zr0.833 | BCC | C | OTHER | -268.8 | 2283.333333333333 |
| Hf0.5Nb0.667Ta0.333Ti1Zr0.833 | BCC | C | OTHER | -196.0 | 1850.0 |
| Hf0.5Nb0.667Ta0.333Ti1Zr0.833 | BCC | C | OTHER | -153.0 | 1520.0 |
| Hf0.5Nb0.667Ta0.333Ti1Zr0.833 | BCC | C | OTHER | -103.0 | 1333.333333333333 |
| Hf0.5Nb0.667Ta0.333Ti1Zr0.833 | BCC | C | OTHER | -43.0 | 1133.333333333333 |
| Hf0.5Nb0.667Ta0.333Ti1Zr0.833 | BCC | C | OTHER | 25.0 | 1046.666666666667 |
| Hf0.5Nb0.667Ta0.333Ti1Zr0.833 | BCC | C | OTHER | 72.0 | 900.0 |
| Hf0.75Nb1Ta0.5Ti1.5Zr1.25 | BCC | C | OTHER | -268.8 | 2230.0 |
| Hf0.75Nb1Ta0.5Ti1.5Zr1.25 | BCC | C | OTHER | -196.0 | 1815.0 |
| Hf0.75Nb1Ta0.5Ti1.5Zr1.25 | BCC | C | OTHER | -153.0 | 1595.0 |
| Hf0.75Nb1Ta0.5Ti1.5Zr1.25 | BCC | C | OTHER | -103.0 | 1375.0 |
| Hf0.75Nb1Ta0.5Ti1.5Zr1.25 | BCC | C | OTHER | -43.0 | 1190.0 |
| Hf0.75Nb1Ta0.5Ti1.5Zr1.25 | BCC | C | OTHER | 25.0 | 1125.0 |
| Hf0.75Nb1Ta0.5Ti1.5Zr1.25 | BCC | C | OTHER | 72.0 | 1030.0 |
| Hf0.75Nb1Ta0.5Ti1.5Zr1.25 | BCC | C | OTHER | -268.8 | 2390.0 |
| Hf0.75Nb1Ta0.5Ti1.5Zr1.25 | BCC | C | OTHER | -196.0 | 1920.0 |
| Hf0.75Nb1Ta0.5Ti1.5Zr1.25 | BCC | C | OTHER | -153.0 | 1370.0 |
| Hf0.75Nb1Ta0.5Ti1.5Zr1.25 | BCC | C | OTHER | -103.0 | 1250.0 |
| Hf0.75Nb1Ta0.5Ti1.5Zr1.25 | BCC | C | OTHER | -43.0 | 1020.0 |
| Hf0.75Nb1Ta0.5Ti1.5Zr1.25 | BCC | C | OTHER | 25.0 | 890.0 |
| Hf0.75Nb1Ta0.5Ti1.5Zr1.25 | BCC | C | OTHER | 72.0 | 640.0 |
| Hf1Mo0.25Nb1Ta1Ti1Zr1 | BCC | C | AC | 25.0 | 1112.0 |
| Hf1Mo0.5Nb1Ta1Ti1Zr1 | BCC | C | AC | 25.0 | 1317.0 |
| Hf1Mo0.5Nb1Ti1V0.5 | BCC | C | AC | 25.0 | 1260.0 |
| Hf1Mo0.5Nb1Ti1V0.5 | BCC | C | AC | 1000.0 | 368.0 |
| Hf1Mo0.5Nb1Ti1V0.5 | BCC | C | AC | 1200.0 | 60.0 |
| Hf1Mo0.75Nb1Ta1Ti1Zr1 | BCC | C | AC | 25.0 | 1373.0 |
| Hf1Mo1Nb1Ta1Ti1 | BCC | C | AC | 25.0 | 1369.0 |
| Hf1Mo1Nb1Ta1Ti1 | BCC | C | AC | 800.0 | 822.0 |
| Hf1Mo1Nb1Ta1Ti1 | BCC | C | AC | 1000.0 | 778.0 |
| Hf1Mo1Nb1Ta1Ti1 | BCC | C | AC | 1200.0 | 699.0 |
| Hf1Mo1Nb1Ta1Ti1 | BCC | C | AC | 1400.0 | 367.0 |
| Hf1Mo1Nb1Ta1Ti1Zr1 | BCC | C | AC | 25.0 | 1512.0 |
| Hf1Mo1Nb1Ta1Ti1Zr1 | BCC | C | AC | 800.0 | 1007.0 |
| Hf1Mo1Nb1Ta1Ti1Zr1 | BCC | C | AC | 1000.0 | 814.0 |
| Hf1Mo1Nb1Ta1Ti1Zr1 | BCC | C | AC | 1200.0 | 556.0 |
| Hf1Mo1Nb1Ta1Zr1 | BCC | C | AC | 25.0 | 1524.0 |
| Hf1Mo1Nb1Ta1Zr1 | BCC | C | AC | 800.0 | 1005.0 |
| Hf1Mo1Nb1Ta1Zr1 | BCC | C | AC | 1000.0 | 927.0 |
| Hf1Mo1Nb1Ta1Zr1 | BCC | C | AC | 1200.0 | 694.0 |
| Hf1Mo1Nb1Ta1Zr1 | BCC | C | AC | 1400.0 | 278.0 |
| Hf1Mo1Nb1Ti1Zr1 | BCC | C | A | 25.0 | 1575.0 |
| Hf1Mo1Nb1Ti1Zr1 | BCC | C | A | 800.0 | 825.0 |
| Hf1Mo1Nb1Ti1Zr1 | BCC | C | A | 1000.0 | 635.0 |
| Hf1Mo1Nb1Ti1Zr1 | BCC | C | A | 1200.0 | 187.0 |
| Hf1Mo1Ta1Ti1Zr1 | BCC | C | AC | 25.0 | 1600.0 |
| Hf1Mo1Ta1Ti1Zr1 | BCC | C | AC | 800.0 | 1045.0 |
| Hf1Mo1Ta1Ti1Zr1 | BCC | C | AC | 1000.0 | 855.0 |
| Hf1Mo1Ta1Ti1Zr1 | BCC | C | AC | 1200.0 | 404.0 |
| Hf1Nb1Ta1Ti1 | BCC | C | A | 25.0 | 847.0 |
| Hf1Nb1Ta1Ti1 | BCC | C | A | 200.0 | 610.0 |
| Hf1Nb1Ta1Ti1 | BCC | C | A | 600.0 | 473.0 |
| Hf1Nb1Ta1Ti1Zr1 | BCC | C | OTHER | 25.0 | 929.0 |
| Hf1Nb1Ta1Ti1Zr1 | BCC | C | OTHER | 400.0 | 790.0 |
| Hf1Nb1Ta1Ti1Zr1 | BCC | C | OTHER | 600.0 | 675.0 |
| Hf1Nb1Ta1Ti1Zr1 | BCC | C | OTHER | 800.0 | 535.0 |
| Hf1Nb1Ta1Ti1Zr1 | BCC | C | OTHER | 1000.0 | 295.0 |
| Hf1Nb1Ta1Ti1Zr1 | BCC | C | OTHER | 1200.0 | 92.0 |
| Hf1Nb1Ta1Zr1 | BCC | C | AC | 25.0 | 1315.0 |
| Hf1Nb1Ti1V1Zr1 | BCC | C | A | 25.0 | 1171.0 |
| Hf1Nb1Ti1Zr1 | BCC | C | AC | 25.0 | 1000.0 |
| Hf1Nb1Ti1Zr1 | BCC | C | AC | 800.0 | 303.0 |
| Hf1Nb1Ti1Zr1 | BCC | C | AC | 1000.0 | 154.0 |
| Mo0.1Nb1Ti1V0.3Zr1 | BCC | C | A | 25.0 | 932.0 |
| Mo0.333Nb0.333Ti0.333V1Zr0.333 | BCC | C | AC | 25.0 | 1418.0 |
| Mo0.3Nb1Ti1V0.3Zr1 | BCC | C | A | 25.0 | 1312.0 |
| Mo0.3Nb1Ti1V1Zr1 | BCC | C | A | 25.0 | 1289.0 |
| Mo0.5Nb0.5Ti0.5V1Zr0.5 | BCC | C | AC | 25.0 | 1538.0 |
| Mo0.5Nb1Ti1V0.3Zr1 | BCC | C | A | 25.0 | 1301.0 |
| Mo0.5Nb1Ti1V1Zr1 | BCC | C | A | 25.0 | 1473.0 |
| Mo0.667Nb0.667Ti0.667V1Zr0.667 | BCC | C | AC | 25.0 | 1735.0 |
| Mo0.7Nb1Ti1V0.3Zr1 | BCC | C | A | 25.0 | 1436.0 |
| Mo0.7Nb1Ti1V1Zr1 | BCC | C | A | 25.0 | 1706.0 |
| Mo1.3Nb1Ti1V0.3Zr1 | BCC | C | AC | 25.0 | 1603.0 |
| Mo1.3Nb1Ti1V1Zr1 | BCC | C | AC | 25.0 | 1496.0 |
| Mo1.5Nb1Ti1V0.3Zr1 | BCC | C | AC | 25.0 | 1576.0 |
| Mo1Nb0.588Ti0.588V0.588Zr0.588 | BCC | C | A | 25.0 | 1645.0 |
| Mo1Nb0.5Ti0.5V0.5Zr0.5 | BCC | C | A | 25.0 | 1765.0 |
| Mo1Nb0.667Ti0.667V0.2Zr0.667 | BCC | C | A | 25.0 | 1576.0 |
| Mo1Nb0.667Ti0.667V0.667Zr0.667 | BCC | C | A | 25.0 | 1603.0 |
| Mo1Nb0.769Ti0.769V0.231Zr0.769 | BCC | C | A | 25.0 | 1603.0 |
| Mo1Nb0.769Ti0.769V0.769Zr0.769 | BCC | C | A | 25.0 | 1496.0 |
| Mo1Nb1Ta1Ti0.25W1 | BCC | C | AC | 25.0 | 1109.0 |
| Mo1Nb1Ta1Ti0.5W1 | BCC | C | AC | 25.0 | 1211.0 |
| Mo1Nb1Ta1Ti0.75W1 | BCC | C | AC | 25.0 | 1304.0 |
| Mo1Nb1Ta1Ti1 | BCC | C | A | 25.0 | 1210.0 |
| Mo1Nb1Ta1Ti1 | BCC | C | A | 60.0 | 1071.0 |
| Mo1Nb1Ta1Ti1 | BCC | C | A | 100.0 | 1029.0 |
| Mo1Nb1Ta1Ti1 | BCC | C | A | 200.0 | 868.0 |
| Mo1Nb1Ta1Ti1 | BCC | C | A | 300.0 | 732.0 |
| Mo1Nb1Ta1Ti1 | BCC | C | A | 400.0 | 685.0 |
| Mo1Nb1Ta1Ti1 | BCC | C | A | 600.0 | 593.0 |
| Mo1Nb1Ta1Ti1 | BCC | C | A | 800.0 | 564.0 |
| Mo1Nb1Ta1Ti1 | BCC | C | A | 1000.0 | 539.0 |
| Mo1Nb1Ta1Ti1V1 | BCC | C | AC | 25.0 | 1400.0 |
| Mo1Nb1Ta1Ti1V1W1 | BCC | C | AC | 25.0 | 1515.0 |
| Mo1Nb1Ta1Ti1V1W1 | BCC | C | AC | 600.0 | 973.0 |
| Mo1Nb1Ta1Ti1V1W1 | BCC | C | AC | 800.0 | 791.3 |
| Mo1Nb1Ta1Ti1V1W1 | BCC | C | AC | 1000.0 | 752.8666666666667 |
| Mo1Nb1Ta1Ti1V1W1 | BCC | C | AC | 1200.0 | 659.0 |
| Mo1Nb1Ta1Ti1W1 | BCC | C | AC | 25.0 | 1380.333333333333 |
| Mo1Nb1Ta1Ti1W1 | BCC | C | AC | 600.0 | 689.0 |
| Mo1Nb1Ta1Ti1W1 | BCC | C | AC | 800.0 | 674.0 |
| Mo1Nb1Ta1Ti1W1 | BCC | C | AC | 1000.0 | 620.0 |
| Mo1Nb1Ta1Ti1W1 | BCC | C | AC | 1200.0 | 586.0 |
| Mo1Nb1Ta1V1 | BCC | C | A | 25.0 | 1525.0 |
| Mo1Nb1Ta1V1W1 | BCC | C | OTHER | 25.0 | 1246.0 |
| Mo1Nb1Ta1V1W1 | BCC | C | OTHER | 600.0 | 862.0 |
| Mo1Nb1Ta1V1W1 | BCC | C | OTHER | 800.0 | 846.0 |
| Mo1Nb1Ta1V1W1 | BCC | C | OTHER | 1000.0 | 842.0 |
| Mo1Nb1Ta1V1W1 | BCC | C | OTHER | 1200.0 | 735.0 |
| Mo1Nb1Ta1V1W1 | BCC | C | OTHER | 1400.0 | 656.0 |
| Mo1Nb1Ta1V1W1 | BCC | C | OTHER | 1600.0 | 477.0 |
| Mo1Nb1Ta1W1 | BCC | C | AC | 25.0 | 1027.0 |
| Mo1Nb1Ta1W1 | BCC | C | AC | 600.0 | 561.0 |
| Mo1Nb1Ta1W1 | BCC | C | AC | 800.0 | 552.0 |
| Mo1Nb1Ta1W1 | BCC | C | AC | 1000.0 | 548.0 |
| Mo1Nb1Ta1W1 | BCC | C | AC | 1200.0 | 506.0 |
| Mo1Nb1Ta1W1 | BCC | C | AC | 1400.0 | 421.0 |
| Mo1Nb1Ta1W1 | BCC | C | AC | 1600.0 | 405.0 |
| Mo1Nb1Ti1 | BCC | C | OTHER | 25.0 | 1100.0 |
| Mo1Nb1Ti1 | BCC | C | OTHER | 1000.0 | 504.0 |
| Mo1Nb1Ti1 | BCC | C | OTHER | 1200.0 | 324.0 |
| Mo1Nb1Ti1V0.25Zr1 | BCC | C | AC | 25.0 | 1776.0 |
| Mo1Nb1Ti1V0.3Zr1 | BCC | C | A | 25.0 | 1455.0 |
| Mo1Nb1Ti1V0.5Zr1 | BCC | C | AC | 25.0 | 1647.0 |
| Mo1Nb1Ti1V0.75Zr1 | BCC | C | AC | 25.0 | 1708.0 |
| Mo1Nb1Ti1V1 | BCC | C | AC | 25.0 | 1200.0 |
| Mo1Nb1Ti1V1Zr1 | BCC | C | A | 25.0 | 1779.0 |
| Mo1Nb1Ti1Zr1 | BCC | C | AC | 25.0 | 1592.0 |
| Mo1Ta1Ti1V1 | BCC | C | AC | 25.0 | 1221.0 |
| Nb0.5Ti0.5V1Zr0.5 | BCC | C | OTHER | 25.0 | 918.0 |
| Nb0.5Ti0.5V1Zr0.5 | BCC | C | OTHER | 600.0 | 571.0 |
| Nb0.5Ti0.5V1Zr0.5 | BCC | C | OTHER | 800.0 | 240.0 |
| Nb0.5Ti0.5V1Zr0.5 | BCC | C | OTHER | 1000.0 | 72.0 |
| Nb1Ta0.3Ti1Zr1 | BCC | C | OTHER | 25.0 | 882.0 |
| Nb1Ta0.3Ti1Zr1 | BCC | C | OTHER | 1000.0 | 274.0 |
| Nb1Ta0.3Ti1Zr1 | BCC | C | OTHER | 1200.0 | 102.0 |
| Nb1Ta1Ti1 | BCC | C | OTHER | 25.0 | 573.0 |
| Nb1Ta1Ti1 | BCC | C | OTHER | 100.0 | 486.0 |
| Nb1Ta1Ti1 | BCC | C | OTHER | 200.0 | 378.0 |
| Nb1Ta1Ti1 | BCC | C | OTHER | 300.0 | 314.0 |
| Nb1Ta1Ti1 | BCC | C | OTHER | 400.0 | 232.0 |
| Nb1Ta1Ti1 | BCC | C | OTHER | 600.0 | 222.0 |
| Nb1Ta1Ti1 | BCC | C | OTHER | 800.0 | 210.0 |
| Nb1Ta1Ti1 | BCC | C | OTHER | 1000.0 | 160.0 |
| Nb1Ta1Ti1V1 | BCC | C | AC | 25.0 | 1028.5 |
| Nb1Ta1Ti1V1W1 | BCC | C | AC | 25.0 | 1420.0 |
| Nb1Ta1Ti1W1 | BCC | C | A | 25.0 | 1054.0 |
| Nb1Ta1Ti1W1 | BCC | C | A | 100.0 | 968.0 |
| Nb1Ta1Ti1W1 | BCC | C | A | 200.0 | 869.0 |
| Nb1Ta1Ti1W1 | BCC | C | A | 300.0 | 754.0 |
| Nb1Ta1Ti1W1 | BCC | C | A | 400.0 | 627.0 |
| Nb1Ta1Ti1W1 | BCC | C | A | 600.0 | 596.0 |
| Nb1Ta1Ti1W1 | BCC | C | A | 800.0 | 564.0 |
| Nb1Ta1Ti1W1 | BCC | C | A | 1000.0 | 459.0 |
| Nb1Ti1V0.3Zr1 | BCC | C | A | 25.0 | 866.0 |
| Nb1Ti1V1Zr1 | BCC | C | OTHER | 25.0 | 1105.0 |
| Nb1Ti1V1Zr1 | BCC | C | OTHER | 600.0 | 834.0 |
| Nb1Ti1V1Zr1 | BCC | C | OTHER | 800.0 | 187.0 |
| Nb1Ti1V1Zr1 | BCC | C | OTHER | 1000.0 | 58.0 |
| Nb1Ti1Zr1 | BCC | C | AC | 25.0 | 1223.0 |
| Nb1Ti1Zr1 | BCC | C | AC | 800.0 | 462.0 |
| Nb1Ti1Zr1 | BCC | C | AC | 1000.0 | 218.0 |

**Table S5.** Dataset comprising fracture strain values (%) for BCC HEAs. The dataset includes columns for alloy composition, phase constitution, mechanical testing method, processing condition, testing temperature (°C), and corresponding yield strength values. In this context, "A" denotes alloys that have undergone Annealing, "AC" represents those in the As Cast condition, and "OTHER" refers to all other processing conditions. The mechanical testing method is specified as "C" for Compression testing.

| **BCC Fracture Strain Dataset** | | | | | |
|---|---|---|---|---|---|
| **Alloy** | **Phases** | **Mechanical Testing** | **Processing Condition** | **Test Temperature C** | **Fracture Strain (%)** |
| Al0.214Nb0.714Ta0.571Ti1V0.143Zr0.929 | BCC | C | OTHER | 25 | 5.0 |
| Al0.214Nb0.714Ta0.714Ti1Zr0.929 | BCC | C | OTHER | 25 | 5.0 |
| Al0.25Co1Cr1Cu1Fe1Mn1Ni1Ti1V1 | BCC | C | AC | 25 | 2.0 |
| Al0.25Mo1Nb1Ti1V1 | BCC | C | AC | 25 | 13.0 |
| Al0.2Mo1Ta1Ti1V1 | BCC | C | AC | 25 | 7.0 |
| Al0.333Nb0.667Ta0.533Ti1V0.133Zr0.667 | BCC | C | OTHER | 25 | 5.0 |
| Al0.3Hf1Nb1Ta1Ti1Zr1 | BCC | C | AC | 25 | 50.0 |
| Al0.3Nb1Ta0.8Ti1.4V0.2Zr1.3 | BCC | C | AC | 25 | 5.0 |
| Al0.3Nb1Ta0.8Ti1.4V0.2Zr1.3 | BCC | C | AC | 800 | 50.0 |
| Al0.3Nb1Ta0.8Ti1.4V0.2Zr1.3 | BCC | C | AC | 1000 | 50.0 |
| Al0.3Nb1Ta1Ti1.4Zr1.3 | BCC | C | AC | 25 | 5.0 |
| Al0.3Nb1Ta1Ti1.4Zr1.3 | BCC | C | AC | 800 | 50.0 |
| Al0.3Nb1Ta1Ti1.4Zr1.3 | BCC | C | AC | 1000 | 50.0 |
| Al0.4Hf0.6Nb1Ta1Ti1Zr1 | BCC | C | AC | 25 | 10.0 |
| Al0.4Hf0.6Nb1Ta1Ti1Zr1 | BCC | C | AC | 800 | 50.0 |
| Al0.4Hf0.6Nb1Ta1Ti1Zr1 | BCC | C | AC | 1000 | 50.0 |
| Al0.5Hf1Nb1Ta1Ti1Zr1 | BCC | C | AC | 25 | 46.0 |
| Al0.5Mo1Nb1Ti1V1 | BCC | C | AC | 25 | 11.0 |
| Al0.5Nb1Ta0.8Ti1.5V0.2Zr1 | BCC | C | AC | 25 | 4.5 |
| Al0.5Nb1Ta0.8Ti1.5V0.2Zr1 | BCC | C | AC | 800 | 50.0 |
| Al0.5Nb1Ta0.8Ti1.5V0.2Zr1 | BCC | C | AC | 1000 | 50.0 |
| Al0.667Nb1Ta0.333Ti1Zr0.333 | BCC | C | OTHER | 25 | 4.0 |
| Al0.667Nb1Ta0.333Ti1Zr0.333 | BCC | C | OTHER | 800 | 12.0 |
| Al0.6Mo1Ta1Ti1V1 | BCC | C | AC | 25 | 4.0 |
| Al0.75Hf1Nb1Ta1Ti1Zr1 | BCC | C | AC | 25 | 30.0 |
| Al0.75Mo1Nb1Ti1V1 | BCC | C | AC | 25 | 8.0 |
| Al1.5Co1Cr1Fe1Ni1Ti1 | BCC | C | AC | 25 | 10.0 |
| Al1Co1Cr1Cu1Ni1Ti1 | BCC | C | AC | 25 | 8.0 |
| Al1Co1Cr1Fe1Mo0.1Ni1 | BCC | C | AC | 25 | 9.0 |
| Al1Co1Cr1Fe1Nb0.1Ni1 | BCC | C | AC | 25 | 17.0 |
| Al1Co1Cr1Fe1Ni1 | BCC | C | AC | 25 | 25.0 |
| Al1Co1Cr1Fe1Ni1Ti0.5 | BCC | C | AC | 25 | 23.0 |
| Al1Co1Cr1Fe1Ni1Ti1 | BCC | C | AC | 25 | 9.0 |
| Al1Cr0.5Nb1Ti1V1 | BCC | C | A | 25 | 0.8 |
| Al1Cr0.5Nb1Ti1V1 | BCC | C | A | 600 | 2.5 |
| Al1Cr1Fe1Mo0.2Ni1 | BCC | C | AC | 25 | 29.0 |
| Al1Cr1Fe1Mo0.5Ni1 | BCC | C | AC | 25 | 13.0 |
| Al1Cr1Fe1Ni1 | BCC | C | AC | 25 | 29.0 |
| Al1Cr1Mo1Nb1Ti1 | BCC | C | AC | 400 | 2.0 |
| Al1Cr1Mo1Nb1Ti1 | BCC | C | AC | 600 | 3.0 |
| Al1Cr1Mo1Nb1Ti1 | BCC | C | AC | 800 | 2.0 |
| Al1Cr1Mo1Nb1Ti1 | BCC | C | AC | 1000 | 15.0 |
| Al1Cr1Mo1Nb1Ti1 | BCC | C | AC | 1200 | 24.0 |
| Al1Mo0.5Nb1Ta0.5Ti1Zr1 | BCC | C | AC | 25 | 10.0 |
| Al1Mo1Nb1Ti1V1 | BCC | C | AC | 25 | 3.0 |
| Al1Nb1.5Ta0.5Ti1.5Zr0.5 | BCC | C | AC | 25 | 4.0 |
| Al1Nb1Ti1V1 | BCC | C | A | 25 | 5.1 |
| Al1Nb1Ti1V1 | BCC | C | A | 600 | 12.25 |
| Al2Co1Cr1Cu1Fe1Mn1Ni1Ti1V1 | BCC | C | AC | 25 | 2.0 |
| Al2Co1Cr1Fe1Ni1Ti1 | BCC | C | AC | 25 | 5.0 |
| Co1Cr1Mo1Nb1Ti0.4 | BCC | C | AC | 25 | 5.0 |
| Cr1Fe1Ni1Ti0.4 | BCC | C | AC | 25 | 6.5 |
| Hf0.24Nb0.23Ti0.38V0.15 | BCC | C | AC | 25 | 20.6 |
| Hf0.5Mo0.5Nb1Ti1Zr1 | BCC | C | AC | 25 | 25.0 |
| Hf0.5Mo1Nb1Ti1Zr1 | BCC | C | AC | 25 | 12.09 |
| Hf0.5Nb0.5Ta0.5Ti1.5Zr1 | BCC | C | AC | 25 | 18.8 |
| Hf1.5Mo1Nb1Ti1Zr1 | BCC | C | AC | 25 | 16.83 |
| Hf1Mo0.25Nb1Ta1Ti1Zr1 | BCC | C | AC | 25 | 50.0 |
| Hf1Mo0.5Nb1Ta1Ti1Zr1 | BCC | C | AC | 25 | 50.0 |
| Hf1Mo0.5Nb1Ti1V0.5 | BCC | C | AC | 25 | 35.0 |
| Hf1Mo0.5Nb1Ti1V0.5 | BCC | C | AC | 1000 | 35.0 |
| Hf1Mo0.5Nb1Ti1V0.5 | BCC | C | AC | 1200 | 35.0 |
| Hf1Mo0.75Nb1Ta1Ti1Zr1 | BCC | C | AC | 25 | 50.0 |
| Hf1Mo1.5Nb1Ti1Zr1 | BCC | C | AC | 25 | 10.83 |
| Hf1Mo1Nb0.5Ti1Zr1 | BCC | C | AC | 25 | 13.02 |
| Hf1Mo1Nb1.5Ti1Zr1 | BCC | C | AC | 25 | 23.97 |
| Hf1Mo1Nb1Ta1Ti1 | BCC | C | AC | 25 | 27.0 |
| Hf1Mo1Nb1Ta1Ti1Zr1 | BCC | C | AC | 25 | 12.0 |
| Hf1Mo1Nb1Ta1Ti1Zr1 | BCC | C | AC | 800 | 23.0 |
| Hf1Mo1Nb1Ta1Ti1Zr1 | BCC | C | AC | 1000 | 30.0 |
| Hf1Mo1Nb1Ta1Ti1Zr1 | BCC | C | AC | 1200 | 30.0 |
| Hf1Mo1Nb1Ti0.5Zr1 | BCC | C | AC | 25 | 12.08 |
| Hf1Mo1Nb1Ti1.5Zr1 | BCC | C | AC | 25 | 28.98 |
| Hf1Mo1Nb1Ti1Zr0.5 | BCC | C | AC | 25 | 18.02 |
| Hf1Mo1Nb1Ti1Zr1 | BCC | C | AC | 25 | 10.12 |
| Hf1Mo1Nb1Ti1Zr1.5 | BCC | C | AC | 25 | 16.09 |
| Hf1Mo1Ta1Ti1Zr1 | BCC | C | AC | 25 | 4.0 |
| Hf1Mo1Ta1Ti1Zr1 | BCC | C | AC | 800 | 19.0 |
| Hf1Mo1Ta1Ti1Zr1 | BCC | C | AC | 1000 | 30.0 |
| Hf1Mo1Ta1Ti1Zr1 | BCC | C | AC | 1200 | 30.0 |
| Hf1Nb1Ta1Ti1Zr1 | BCC | C | AC | 25 | 50.0 |
| Hf1Nb1Ta1Zr1 | BCC | C | AC | 25 | 21.6 |
| Hf1Nb1Ti1V1Zr1 | BCC | C | AC | 25 | 29.6 |
| Hf1Nb1Ti1Zr1 | BCC | C | AC | 25 | 52.0 |
| Hf1Nb1Ti1Zr1 | BCC | C | AC | 800 | 51.0 |
| Hf1Nb1Ti1Zr1 | BCC | C | AC | 1000 | 51.0 |
| Mo0.1Nb1Ti1V0.3 | BCC | C | AC | 25 | 45.0 |
| Mo0.1Nb1Ti1V0.3Zr1 | BCC | C | AC | 25 | 45.0 |
| Mo0.333Nb0.333Ti0.333V1Zr0.333 | BCC | C | AC | 25 | 24.0 |
| Mo0.3Nb1Ti1V0.3 | BCC | C | AC | 25 | 50.0 |
| Mo0.3Nb1Ti1V0.3Zr1 | BCC | C | AC | 25 | 50.0 |
| Mo0.3Nb1Ti1V1Zr1 | BCC | C | AC | 25 | 42.0 |
| Mo0.5Nb0.5Ti0.5V1Zr0.5 | BCC | C | AC | 25 | 23.0 |
| Mo0.5Nb1Ti1V0.3 | BCC | C | AC | 25 | 43.0 |
| Mo0.5Nb1Ti1V0.3Zr1 | BCC | C | AC | 25 | 43.0 |
| Mo0.5Nb1Ti1V1Zr1 | BCC | C | AC | 25 | 32.0 |
| Mo0.667Nb0.667Ti0.667V1Zr0.667 | BCC | C | AC | 25 | 20.0 |
| Mo0.7Nb1Ti1V0.3 | BCC | C | AC | 25 | 27.0 |
| Mo0.7Nb1Ti1V0.3Zr1 | BCC | C | AC | 25 | 26.6 |
| Mo0.7Nb1Ti1V1Zr1 | BCC | C | AC | 25 | 32.0 |
| Mo1.0Nb1Ti1V0.3Zr1 | BCC | C | AC | 25 | 25.0 |
| Mo1.3Nb1Ti1V0.3 | BCC | C | AC | 25 | 20.0 |
| Mo1.3Nb1Ti1V0.3Zr1 | BCC | C | AC | 25 | 20.0 |
| Mo1.3Nb1Ti1V1Zr1 | BCC | C | AC | 25 | 30.0 |
| Mo1.3Nb1Ti1V1Zr1 | BCC | C | AC | 25 | 30.0 |
| Mo1.5Nb1Ti1V0.3 | BCC | C | AC | 25 | 8.0 |
| Mo1.5Nb1Ti1V0.3Zr1 | BCC | C | AC | 25 | 8.0 |
| Mo1.5Nb1Ti1V1Zr1 | BCC | C | AC | 25 | 20.0 |
| Mo1.7Nb1Ti1V1Zr1 | BCC | C | AC | 25 | 15.0 |
| Mo1Nb1Ta1Ti0.25W1 | BCC | C | AC | 25 | 2.5 |
| Mo1Nb1Ta1Ti0.5W1 | BCC | C | AC | 25 | 5.9 |
| Mo1Nb1Ta1Ti0.75W1 | BCC | C | AC | 25 | 8.4 |
| Mo1Nb1Ta1Ti1V1 | BCC | C | AC | 25 | 30.0 |
| Mo1Nb1Ta1Ti1V1W1 | BCC | C | AC | 25 | 10.6 |
| Mo1Nb1Ta1Ti1W1 | BCC | C | AC | 25 | 14.1 |
| Mo1Nb1Ta1V1 | BCC | C | A | 25 | 21.0 |
| Mo1Nb1Ta1V1W1 | BCC | C | AC | 25 | 2.0 |
| Mo1Nb1Ta1V1W1 | BCC | C | AC | 600 | 13.0 |
| Mo1Nb1Ta1V1W1 | BCC | C | AC | 800 | 17.0 |
| Mo1Nb1Ta1V1W1 | BCC | C | AC | 1000 | 19.0 |
| Mo1Nb1Ta1V1W1 | BCC | C | AC | 1200 | 7.5 |
| Mo1Nb1Ta1V1W1 | BCC | C | AC | 1400 | 40.0 |
| Mo1Nb1Ta1V1W1 | BCC | C | AC | 1600 | 40.0 |
| Mo1Nb1Ta1W1 | BCC | C | AC | 25 | 2.0 |
| Mo1Nb1Ta1W1 | BCC | C | AC | 600 | 40.0 |
| Mo1Nb1Ta1W1 | BCC | C | AC | 800 | 40.0 |
| Mo1Nb1Ta1W1 | BCC | C | AC | 1000 | 40.0 |
| Mo1Nb1Ta1W1 | BCC | C | AC | 1200 | 40.0 |
| Mo1Nb1Ta1W1 | BCC | C | AC | 1400 | 40.0 |
| Mo1Nb1Ta1W1 | BCC | C | AC | 1600 | 40.0 |
| Mo1Nb1Ta1W1 | BCC | C | OTHER | 25 | 7.5 |
| Mo1Nb1Ti1V0.25Zr1 | BCC | C | AC | 25 | 30.0 |
| Mo1Nb1Ti1V0.3 | BCC | C | AC | 25 | 25.0 |
| Mo1Nb1Ti1V0.5Zr1 | BCC | C | AC | 25 | 28.0 |
| Mo1Nb1Ti1V0.75Zr1 | BCC | C | AC | 25 | 29.0 |
| Mo1Nb1Ti1V1 | BCC | C | AC | 25 | 26.0 |
| Mo1Nb1Ti1V1.5Zr1 | BCC | C | AC | 25 | 20.0 |
| Mo1Nb1Ti1V1Zr1 | BCC | C | AC | 25 | 26.0 |
| Mo1Nb1Ti1V2Zr1 | BCC | C | AC | 25 | 23.0 |
| Mo1Nb1Ti1V3Zr1 | BCC | C | AC | 25 | 24.0 |
| Mo1Nb1Ti1Zr1 | BCC | C | AC | 25 | 34.0 |
| Mo1Ta1Ti1V1 | BCC | C | AC | 25 | 10.0 |
| Mo2Nb1Ti1V1Zr1 | BCC | C | AC | 25 | 12.0 |
| Nb1Ta1Ti1V1 | BCC | C | OTHER | 25 | 28.566666666666667 |
| Nb1Ta1Ti1V1W1 | BCC | C | AC | 25 | 20.0 |
| Nb1Ta1V1W1 | BCC | C | AC | 25 | 12.0 |
| Nb1Ta1V1W1 | BCC | C | AC | 25 | 12.0 |
| Nb1Ti1V0.3Zr1 | BCC | C | AC | 25 | 45.0 |
| Nb1Ti1V1Zr1 | BCC | C | AC | 25 | 50.0 |
| Nb1Ti1V2Zr1 | BCC | C | AC | 25 | 50.0 |
| Nb1Ti1Zr1 | BCC | C | AC | 25 | 48.0 |
| Nb1Ti1Zr1 | BCC | C | AC | 800 | 51.0 |
| Nb1Ti1Zr1 | BCC | C | AC | 1000 | 51.0 |